\documentclass[%
 reprint,
 amsmath,amssymb,
 aps,
floatfix
]{revtex4-2}
\usepackage{graphicx}% Include figure files
\usepackage{dcolumn}% Align table columns on decimal point
\usepackage{bm}% bold math
\usepackage{appendix}
\usepackage{comment}

\begin{document}

\title{A statistical model for quantum spin and photon number states}

\author{Sam Powers, Guangpeng Xu, Herbert Fotso, Tim Thomay, and Dejan Stojkovic}%
\email{sampower@buffalo.edu or ds77@buffalo.edu}
\affiliation{Department of Physics SUNY at Buffalo, Buffalo, NY 14260-1500, USA}%

\date{\today}% It is always \today, today,

\begin{abstract}
The most irreducible way to represent information is a sequence of two symbols. In this paper, we construct quantum states using this basic building block. Specifically, we show that the probabilities that arise in quantum theory can be reduced to counting more fundamental ontic states, which we interpret as event networks and model using sequences of 0's and 1's. A completely self contained formalism is developed for the purpose of organizing and counting these ontic states, which employs the finite cyclic group $\mathbb{Z}_2 = \{0, 1\}$, basic set theory, and combinatorics. This formalism is then used to calculate probability distributions associated with particles of arbitrary spin interacting with sequences of two rotated Stern-Gerlach detectors. A central ingredient of this construction is the rule which converts the abstract counts labelling an operation into the physical angle of rotation it represents. We show that this rule is not linear in the counts, but is instead fixed by the half-angle law $\tan(\theta_{ab}/2)=\tilde{B}_{map}/\tilde{A}_{map}$, which is the unique assignment consistent with the composition of successive rotations. These calculations are compared with the predictions of non-relativistic quantum mechanics and shown to agree exactly in the limit of large sequence length $n$, with finite $n$ corrections which vanish as $O(1/n)$ and with no free or fitted parameters. The residual deviation at finite $n$ does not lead to violations of relevant no-go theorems, such as Bell's inequalities, the Kochen-Specker theorem, or the PBR theorem. The proposed model is then extended to an optical system involving photon number states passing through a beam splitter. Leveraging recent advancements in high precision experiments on these systems, we then propose a means of testing the new model using a tabletop experiment.

\end{abstract}

\maketitle

\section{Introduction\label{sec:Introduction}}

In 1922, physicists Otto Stern and Walther Gerlach reported their experimental results concerning the discrete nature of angular momentum \citep{Bauer2023}. 
The experiment they performed was proposed as a test of the Bohr-Sommerfeld model of the atom, where electrons were restricted to fixed orbits around the nucleus, a feature known then as ``space quantization'' \citep{SchmidtBoecking2016}. Their experiment showed that the projection of angular momentum was indeed quantized along the spatial axis of their choice. The Bohr-Sommerfeld model was later supplanted, but the experimental results of Stern and Gerlach (SG) continued to play an important role in the development of modern quantum mechanics (QM). Today, sequences of SG detectors are of significant pedagogical importance, often being introduced in the early chapters of undergraduate texts on QM \citep{Sakurai19941995}.

\begin{figure}
\begin{centering}
\includegraphics[scale=.6]{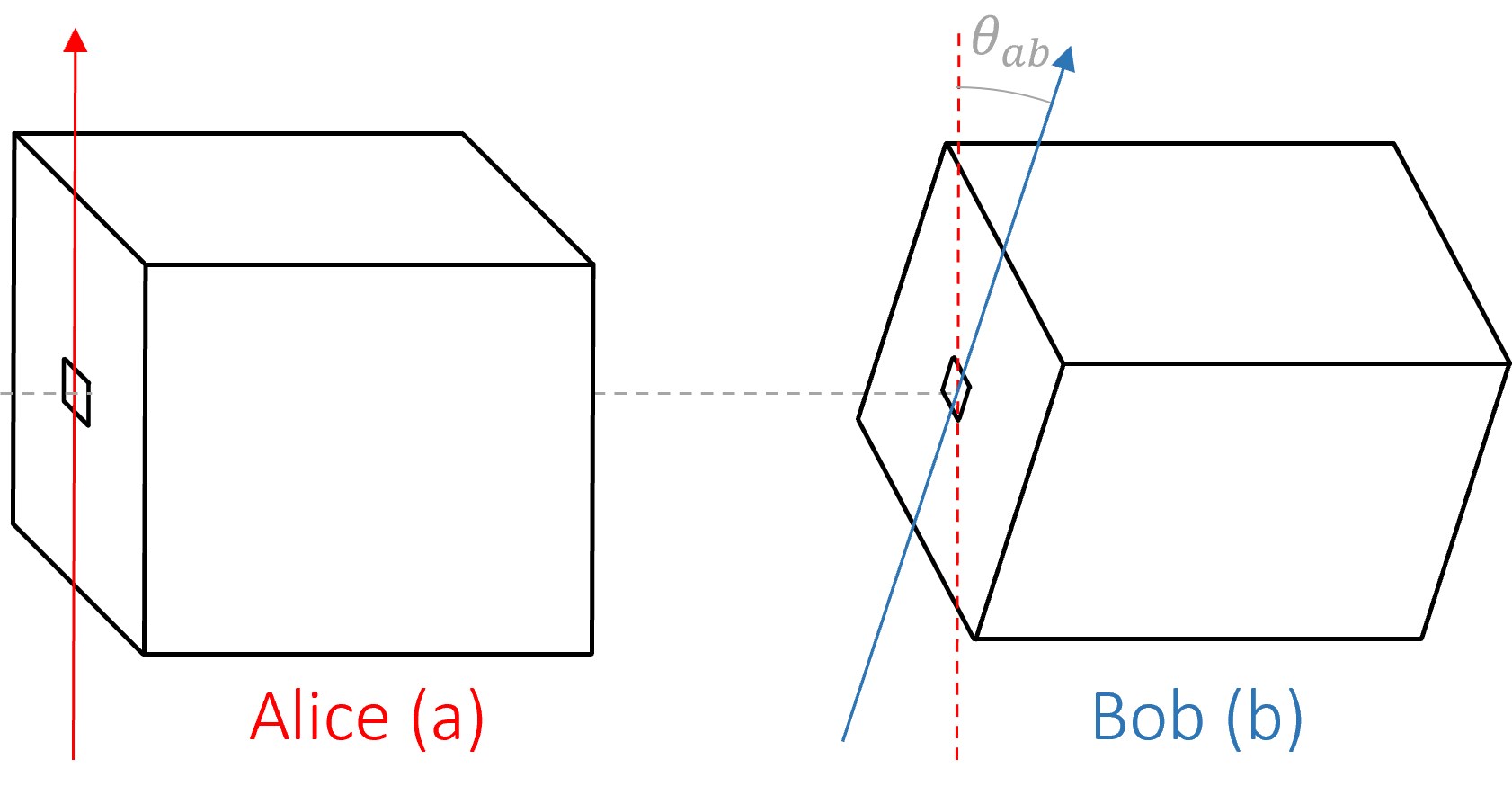}
\par\end{centering}
\caption{Within each SG detector is a magnetic field gradient, the spatial orientation of which is indicated by a red arrow for Alice's detector and a blue arrow for Bob's detector.  Incoming particles are deflected by these field gradients at an angle which depends on their spin projection quantum number $m$ within the chosen spatial frame. The rotations modeled here are limited to $\theta_{ab}$, which is defined in the plane perpendicular to the beam axis.}\label{fig: sketch 1}
\end{figure}

Wrapped up in the treatment of sequences of SG detectors are many
foundational questions in physics. There are the familiar ones surrounding
QM, such as non-determinism, non-commutativity, and the measurement
problem. There are also questions about the nature of spin, the intrinsic
angular momentum carried by fundamental particles. In particular,
what is the origin of this degree of freedom, why is it quantized,
and what is its relationship to space and time? We raise these questions to highlight the non-trivial physics involved in experiments comprised of sequences of SG detectors, making them an excellent testing ground for alternative models. Our primary focus will be a new model for experiments involving two SG detectors that may differ in spatial orientation. A depiction of this experimental setup is provided in Figure \ref{fig: sketch 1}, where the rotations being modeled are limited to the angle $\theta_{ab}$. 

In this paper, we show that the probability distributions generated by the canonical treatment of this physical system can be reproduced by  counting ontic states. Ontic state spaces are constructed from first principles, with no presuppositions about the modeled physical system, other than it should be possible to represent using 0's and 1's alone. Indeed, the work presented herein may be viewed as a small step towards Wheeler's dream of ``\textit{it from bit}":

\begin{quote}
    Otherwise stated, every physical quantity, every it, derives its ultimate significance from bits, binary yes-or-no indications, a conclusion which we epitomize in the phrase, \textit{it from bit} \citep{Wheeler1989}
\end{quote}

Within the literature on quantum foundations, the proposed model most closely relates to the statistical interpretation of canonical quantum theory, where the quantum state is an ensemble of more fundamental, but indistinguishable ontic states \citep{Ballentine1970}. Here, sequences of 0's and 1's are used to model these ontic states, where the quantum state is encoded in the number of 0's and 1's. Thus, the information necessary to distinguish two ontic states belonging to the same quantum state is stored in the position of these symbols within the sequence, which is hidden from observers. While no ontic state is more likely than any other, some quantum states have more ontic states than others, making them more likely to occur. The formalism developed herein can be used to organize and count these ``hidden" ontic states, enabling the calculation of relative frequencies, or probability distributions. 

Typically, hidden information approaches to quantum theory attempt to recover classical notions like local realism. However, no-go theorems like Bell's theorem and the Kochen-Specker theorem rule out these classical ontologies \citep{Bell1964, Kochen1968}, under the assumption of statistical independence \citep{Hossenfelder2020}. Of course, these theorems do not rule out hidden information models in general. Rather, they dictate the form these models can take. While a definitive physical interpretation of the underlying ontic states and their base-2 building blocks remains an open issue, the ``event network" picture has proven to be a sufficient conceptual basis for the systems being modeled here. This may be viewed as an alternative to the wave-particle picture of canonical quantum theory. Though the choice to interpret ontic states as event networks is rooted in pragmatism, we are also motivated by empiricism and epistemology. After all, the measurement event is our lone source of information about the physical systems we intend to model. It seems only natural that it play a central role in our model of that physical system. 

The event networks of interest here consist of two measurement events (nodes) separated by some operation (edge). Measurement events are modeled by ordered pairs of base-2 sequences, where one is always associated with the observer participating in that measurement event. Encoded in the relationship between these base-2 sequences are the total spin ($j$) and the spin projection ($m$) quantum numbers. The intervening operation is also modeled by an ordered pair of base-2 sequences, which are equipped with the addition modulo two operator ($\oplus$). Encoded in this operation is the relative angle of rotation ($\theta_{ab}$) between the two Stern-Gerlach detectors depicted in Figure \ref{fig: sketch 1}.

The ``event-centric" nature of the proposed model suggests that insight into the broader physical picture implied by this work may be found by studying event-centric formulations of canonical quantum theory. Perhaps the most widely regarded of these are ``Consistent Histories", introduced by Robert Griffiths in 1984  \citep{Griffiths1984,Griffiths2002}, and the closely related ``Decoherent Histories", proposed by Murray Gell-Mann and James B. Hartle several years later \citep{gellmann2018}. In the Consistent Histories formulation of quantum theory, projection operators of a Hilbert Space, or ``projectors", are interpreted as events. The set of all possible events associated with a given physical system constitute a sample space upon which an event algebra is defined. A history can then be constructed from the elements of this algebra, which can be assigned a probability provided a consistency condition related to state orthogonality is met. More recent work concerning applications of Consistent Histories to conceptual issues in canonical quantum theory can be found in \citep{Griffiths2020} and \citep{Griffiths2023}. Other notable efforts to reformulate quantum theory in terms of events include ``Quantum Measure Theory" \citep{Sorkin1994, Sorkin1997, Dowker2023, Chakraborti2024}, ``Events, Trees, Histories" (ETH) \citep{froehlich2019} and ``Geometric Event-Based Quantum Mechanics" (GEB) \citep{Lloyd2023}, all of which constitute exciting research programs in their own right.

The discrete nature of this formalism further differentiates the proposed model from much of the literature concerning the foundations of physics. This is due primarily to the ubiquitous assumption of continuity in theoretical physics. Importantly, this assumption can never be empirically confirmed due to the discrete nature of empirical data. That is, the continuity of the optical spectrum, for example, is not a testable prediction of classical or quantum optics. Challenging this assumption was a central motivation for the work presented here. As the length ($n$) of the sequences used to model ontic states becomes large, the number of possible observable outcomes grows while probabilities become smooth. In the limit that $n$ goes to $\aleph_0$, or the countable infinity, the cardinality of ontic state spaces generally go to $\aleph_1$, or the uncountable infinity. It is in this limit that a correspondence to canonical quantum theory may be found.

Over the past six decades, there have been several notable attempts to challenge various manifestations of the continuity assumption. The most prominent of these attempts being Roger Penrose's work on spin networks and the subsequent development of Loop Quantum Gravity \citep{Penrose1971, Ashtekar2021}. This has spawned many interesting ideas, including the closely related project of ``Polymer Quantum Mechanics" \citep{Ashtekar2003,Corichi2007}. More recently, the Cellular Automata program has gained some attention through the work of Gerard 't Hooft \citep{Hooft2014} and Stephen Wolfram \citep{Wolfram2002}. Also of relevance is the work of David Finkelstein on the issue of a Space-Time Code \citep{Finkelstein1969}, as well as the work by Chang et al. on their Galois Field Quantum Mechanics \citep{CHANG2013.1, Chang2013.2, Chang2014, Chang2019}. Though not directly related to the issue of continuity, the Amplituhedron research program initiated by Arkani-Hamed et al. is relevant here due to its use of combinatorial techniques in the calculation of probabilities \citep{Arkani_Hamed2014, Arkani_Hamed2018}. Collectively, these research efforts have contributed significantly to the prior art upon which the present work is based.

The body of this paper begins in section \ref{sec: Formalism} with a bottom-up construction of the proposed formalism, culminating with our expression for the probabilities of interest. This is followed by an extended pedagogical development of the new model. We begin with a study of the combinatorial terms appearing in the canonical treatment of the modeled system. Using the conclusions of this section, we then embark on an incremental approach to model development, starting with the special case of a spin $\frac{1}{2}$ particle. This incremental process culminates in section \ref{sec: Spin 1/2 arbitrary n}, where the general expression for probability distributions associated with arbitrary spin systems and rotations is constructed. We then address the issue of interference in section \ref{sec: Spin 1 arbitrary n}, illustrated by the treatment of a spin $1$ system. In section \ref{sec:Testing-the-model}, we show how the new model can be applied to an optical system, which provides an experimental advantage over spin systems. We also discuss several ways to test the proposed model. Finally, in section \ref{sec: Discussion}, we offer a short discussion regarding the position of this new model within the quantum foundations literature, as well as plans for future work.

\section{Formalism \label{sec: Formalism}}

The building blocks of this formalism are the symbols $0$ and $1$, equipped with addition modulo two ($\oplus$) and the Cartesian product ($\otimes$).\\

Let $S^1(n)$ be the set of all $n^{th}$ order Cartesian products of $0$'s and $1$'s, where $n\in \mathbb{N}_0$. The elements of this set will be called base-2 sequences herein. \\

Let $\tilde{0},\tilde{1}\in \mathbb{N}_0$ be the number of $0$'s and $1$'s that appear in a given base-2 sequence, where $n=\tilde{0}+\tilde{1}$. Symbols with a tilde on top will be called counts herein. \\

Let $s^1 \in S^1(\tilde{0},\tilde{1})$ be a single base-2 sequence with $\tilde{0}$ $0$'s and $\tilde{1}$ $1$'s. The number of base-2 sequences in the set $S^1(\tilde{0},\tilde{1})$, or its cardinality, can be expressed using factorial notation:

\begin{equation}
    |S^1(\tilde{0},\tilde{1})| = \frac{n!}{\tilde{0}!\tilde{1}!}
\end{equation}

Let $s^1_1 \in S_1^1(\tilde{0}_1,\tilde{1}_1)$, $s^1_2 \in  S_2^1(\tilde{0}_2,\tilde{1}_2)$, $s^1_3 \in S_3^1(\tilde{0}_3,\tilde{1}_3)$ be indexed base-2 sequences of the same length:

\begin{equation}
    \tilde{0}_1 + \tilde{1}_1 = \tilde{0}_2 + \tilde{1}_2 = \tilde{0}_3 + \tilde{1}_3 = n
\end{equation}

Element-wise addition modulo two can be used to map indexed base-2 sequences to one another, where $d_{12}$ is the number of differences between $s^1_1$ and $s^1_2$, or their Hamming distance, and all subscripts are interchangeable:
\begin{equation}
\begin{aligned}
    s^1_1 \oplus &s^1_2 = s^1_3 \\
    d_{12}&=\tilde{1}_3
\end{aligned}
\end{equation}

Element-wise Cartesian products can be used to construct base-4 sequences from two base-2 sequences like so, where lowercase Latin subscripts label sequences associated with observers in the physical model of interest:

\begin{equation}
    s^1_a\otimes s^1_1 \equiv s^2_{a}
\end{equation}\\

The base-4 sequence $s^2_{a}$ is thus an ordered pair, one member of which is Alice's reference sequence $s^1_a$ and the other of which carries the numeral index. Since the numeral is fixed once the observer is named, we suppress it throughout and label base-4 sequences, their counts and their quantum numbers by the observer alone, writing $s^2_a$, $\tilde{C}_a$, $m_a$ and so on in place of $s^2_{a1}$, $\tilde{C}_{a1}$ and $m_{a1}$. The only exception is section \ref{sec: Spin 1/2 n=2}, where the ordering within the pair is itself the subject and the numeral is displayed.

\begin{figure}
\begin{centering}
\includegraphics[width=\columnwidth]{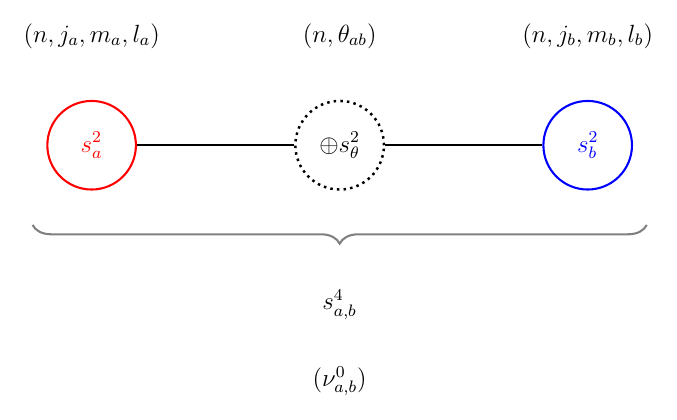}
\par\end{centering}
\caption{A diagram of the event network for a spin system interacting with two SG detectors, as illustrated in Figure \ref{fig: sketch 1}. The measurement events occurring in Alice's (red node) and Bob's (blue node) detectors are modeled by base-4 sequences and labeled with four base-4 quantum numbers. The rotation operation (black edge) is modeled using a base-4 sequence equipped with the $\oplus$ operator and labeled with two base-4 quantum numbers. The full network is modeled by a base-16 sequence, which has one additional degree of freedom not included in the base-4 quantum numbers.}\label{fig: base-16 network}
\end{figure}

Base-4 sequences that include a Latin subscript are used to model measurement events, or the nodes within the networks of interest (Figure \ref{fig: base-16 network}). The symbols comprising this base-4 sequence are defined as follows: 

\begin{equation}
    \begin{aligned}
        A_{a}\equiv 0_a\otimes 0_1, \quad
        B_{a}\equiv 1_a\otimes 1_1\\
        C_{a}\equiv 1_a\otimes 0_1, \quad
        D_{a}\equiv 0_a\otimes 1_1
    \end{aligned}
    \label{eq: base-4 definitions}
\end{equation}

The base-4 counts $\tilde{A}_{a},\tilde{B}_{a},\tilde{C}_{a},\tilde{D}_{a}\in \mathbb{N}_0$ are used to define the quantum numbers that label the measurement events in the proposed physical model, where $d_{a}$ is the Hamming distance between $s^1_a$ and $s^1_1$:

\begin{equation}
    \begin{aligned}
        n & = \tilde{A}_{a}+\tilde{B}_{a}+\tilde{C}_{a}+\tilde{D}_{a}\\
        j_{a}&\equiv \frac{1}{2}(\tilde{C}_{a}+\tilde{D}_{a})=\frac{1}{2}d_{a}\\
        m_{a}& \equiv \frac{1}{2}(\tilde{C}_{a}-\tilde{D}_{a})\\
        l_{a}&\equiv \frac{1}{2}(\tilde{A}_{a}-\tilde{B}_{a})
        \label{eq: base-4 quantum numbers}
    \end{aligned}
\end{equation}

The definitions of the spin quantum number (j) and spin projection quantum number (m) are motivated by their relationship to the Hamming distance \citep{Powers2022}. The physical interpretation of $l_{a}$ has yet to be determined and may therefore be defined in other ways. For any choice of these base-4 quantum numbers, along with the inverted relations in Table \ref{tab: base-4}, the cardinality of the associated set of base-4 sequences $S^2_{a}(n,j_{a},m_{a},l_{a})$ can be calculated like so:

\begin{equation}
    |S^2_{a}(n,j_{a},m_{a},l_{a})|=\frac{n!}{\tilde{A}_{a}!\tilde{B}_{a}!\tilde{C}_{a}!\tilde{D}_{a}!}
\end{equation}

Element-wise addition modulo two can be used to map any base-4 sequence to another, where $d_{a,b}$ is the Hamming distance between $s^2_{a}$ and $s^2_{b}$:

\begin{equation}
    \begin{aligned}
        s^2_{a}\oplus s^2_{map}& = s^2_{b}\\
        d_{a,b}=\tilde{B}_{map}+&\tilde{C}_{map}+\tilde{D}_{map}
    \end{aligned}
\end{equation}

For the model of interest here, we enforce the constraint $j_{a}=j_{b}$, which is accomplished by setting $\tilde{C}_{map}=\tilde{D}_{map}=0$. The quantities $j_{a}$ and $j_{b}$ will now be referred to collectively as $j$. The two surviving map counts then obey $\tilde{A}_{map}+\tilde{B}_{map}=n$, and the remaining degrees of freedom for $s^2_{map}$ are used to model the angle of rotation ($\theta_{ab}$) between Alice's and Bob's detectors:

\begin{equation}
    \begin{aligned}
        \tan\left(\frac{\theta_{ab}}{2}\right)=&\frac{\tilde{B}_{map}}{\tilde{A}_{map}}\\
        s^2_{map}&\rightarrow s^2_{\theta}
    \end{aligned}
    \label{eq: calibration}
\end{equation}

The rule in equation (\ref{eq: calibration}) is not a matter of convention, and it is established in section \ref{sec: Spin 1/2 n=1}. As $\tilde{B}_{map}$ runs over the integers $0,1,\dots,n$ it sweeps $\theta_{ab}$ over the full range $[0,\pi]$.

The operator $\oplus s^2_{\theta}$ is used to model the evolution from one measurement event to another, or the edge in the networks of interest (Figure \ref{fig: base-16 network}). Element-wise Cartesian products can be used to construct base-16 sequences from two base-4 sequences like so:

\begin{equation}
    s^2_{a}\otimes s^2_{b} \equiv s^4_{a,b}
\end{equation}

Base-16 sequences are used to model a network of two measurement events, or two nodes connected by an edge. Due to the constraint $\tilde{C}_{map}=\tilde{D}_{map}=0$, only eight of the sixteen base-16 counts can have non-zero values for the model of interest:

\begin{equation}
    \begin{aligned}
        \widetilde{AA},\widetilde{AB},\widetilde{BA},\widetilde{BB},\widetilde{CC},\widetilde{CD},\widetilde{DC},\widetilde{DD} &\in \mathbb{N}_0\\
    \widetilde{AC},\widetilde{AD},\widetilde{BC},\widetilde{BD},\widetilde{CA},\widetilde{CB},\widetilde{DA},\widetilde{DB} &= 0
    \end{aligned}
\end{equation}

The seven quantum numbers ($n$, $j$, $m_{a}$, $m_{b}$, $l_{a}$, $l_{b}$, $\theta_{ab}$) previously defined in terms of base-4 counts can be expressed in terms of these base-16 counts:

\begin{equation}
n=\widetilde{AA}+\widetilde{AB}+\widetilde{BA}+\widetilde{BB}+\widetilde{CC}+\widetilde{CD}+\widetilde{DC}+\widetilde{DD}\label{eq: n}
\end{equation}

\begin{equation}
\begin{aligned}
    j=&\frac{1}{2}(\widetilde{CC}+\widetilde{CD}+\widetilde{DC}+\widetilde{DD})\\
    m_{a}=&\frac{1}{2}(\widetilde{CC}+\widetilde{CD}-\widetilde{DC}-\widetilde{DD}) \\ m_{b}=&\frac{1}{2}(\widetilde{CC}+\widetilde{DC}-\widetilde{CD}-\widetilde{DD})\\
    l_{a}=&\frac{1}{2}(\widetilde{AA}+\widetilde{AB}-\widetilde{BA}-\widetilde{BB}) \\ l_{b}=&\frac{1}{2}(\widetilde{AA}+\widetilde{BA}-\widetilde{AB}-\widetilde{BB})
\end{aligned}
\end{equation}

\begin{equation}
\tan\left(\frac{\theta_{ab}}{2}\right)=\frac{\widetilde{AB}+\widetilde{BA}+\widetilde{CD}+\widetilde{DC}}{\widetilde{AA}+\widetilde{BB}+\widetilde{CC}+\widetilde{DD}}
\label{eq: theta base-8}
\end{equation}

In the networks of interest, the quantum numbers ($n$, $j$, $m_{a}$, $m_{b}$, $l_{a}$, $l_{b}$) are used to label measurement events (nodes), while ($n$, $\theta_{ab}$) are used to label the intervening operation (edge). The eighth and final quantum number is not a property of measurement events (nodes), nor the intervening operation (edge). Rather, it is an emergent property of the complete network (Figure \ref{fig: base-16 network}), where the superscript indicates that this quantity is part of a large family of related quantities:

\begin{equation}
    \nu^0_{a,b}\equiv\frac{1}{4}(\widetilde{CD}+\widetilde{DC}+\widetilde{AA}+\widetilde{BB})
\end{equation}

For the physical model of interest, these eight variables can be categorized as random $R=(m_{b})$, conditional $U=(n,j,m_{a},\theta_{ab})$, nuisance $W=(l_{a},l_{b})$, and path $\Lambda=(\nu^0_{a,b})$, where $r$, $u$, $w$, and $\lambda$ are the corresponding N-tuples. The complete set of quantum numbers is then denoted as $Q=R\cup U\cup W \cup \Lambda$, which can be found for any combination of $r$, $u$, $w$, and $\lambda$ by solving linear Diophantine equations. With this shorthand, along with the inverted relations in Table \ref{tab: base-8}, the cardinality of the set $S^4_{a,b}(r,u,w,\lambda)$ can be calculated like so:

\begin{equation}
    |S^4_{a,b}(r,u,w,\lambda)|=\frac{n!}{\widetilde{AA}!\widetilde{AB}!\cdots \widetilde{DD}!}
    \label{eq: base-4 cardinality}
\end{equation}

Using the formalism introduced thus far, a counting procedure can be constructed that generates the probability distributions of interest. The building blocks of this construction are the contextual sets $\varepsilon_a(r,u,w,\lambda)$ and $\varepsilon_b(r,u,w,\lambda)$:

\begin{equation}
    \begin{aligned}
        \varepsilon_a(r,u,w,\lambda) \equiv \{s^4_{a,b}|s^2_{a},r,u,w,\lambda\}\\
        \varepsilon_b(r,u,w,\lambda) \equiv \{s^4_{a,b}|s^2_{b},r,u,w,\lambda\}
        \label{eq: contextual sets}
    \end{aligned}
\end{equation}

The contextual set $\varepsilon_a(r,u,w,\lambda)$ contains all of the base-16 sequences $s^4_{a,b}$ compatible with the quantum numbers $(r,u,w,\lambda)$ and a fixed choice of $s^2_{a}$. Likewise, $\varepsilon_b(r,u,w,\lambda)$ contains all of the base-16 sequences $s^4_{a,b}$ compatible with the quantum numbers $(r,u,w,\lambda)$ and a fixed choice of $s^2_{b}$. The cardinality of these sets can be calculated like so, where $|\varepsilon_a|,|\varepsilon_b|\in \mathbb{N}_0$:

\begin{equation}
    \begin{aligned}
        |\varepsilon_a(r,u,w,\lambda)| &= \frac{|S^4_{a,b}(r,u,w,\lambda)|}{|S^2_{a}(n,j,m_{a},l_{a})|}  \\
        |\varepsilon_b(r,u,w,\lambda)| &= \frac{|S^4_{a,b}(r,u,w,\lambda)|}{|S^2_{b}(n,j,m_{b},l_{b})|}
    \end{aligned}
\end{equation}

Interference, driven by the deviation of the path quantum number $\lambda$ from some arbitrary reference value $\lambda_0$, is accounted for in alternating sums over these contextual set cardinalities, where $\Upsilon_a,\Upsilon_b \in \mathbb{Z}$ and always share the same sign:

\begin{equation}
    \begin{aligned}
        \Upsilon_a(r,u,w)\equiv \sum_{\lambda \in Q(r,u,w)}(-1)&^{\Delta \lambda(\lambda,\lambda_0)}|\varepsilon_a(r,u,w,\lambda)|\\
        \Upsilon_b(r,u,w)\equiv \sum_{\lambda \in Q(r,u,w)}(-1)&^{\Delta \lambda(\lambda,\lambda_0)}|\varepsilon_b(r,u,w,\lambda)|\\
        \Delta \lambda(\lambda,\lambda_0) \equiv&  \sum_{i}\lambda^i - \lambda^i_0
    \end{aligned}
\end{equation}

\begin{figure}
\begin{centering}
\includegraphics[scale=1]{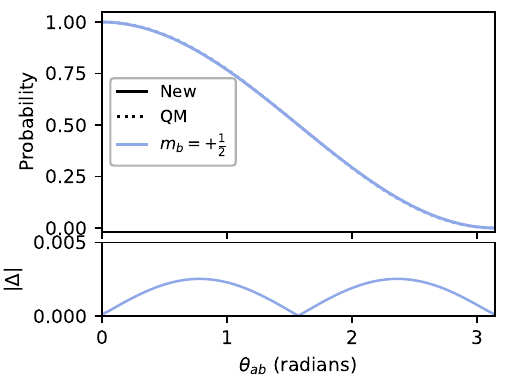}
\hspace{1cm}
\includegraphics[scale=1]{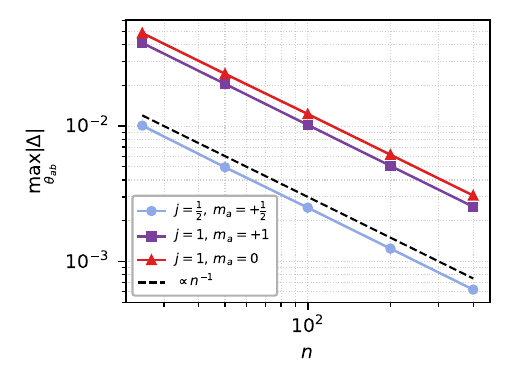}
\par\end{centering}
\caption{A comparison of models for $n=100$, $j=\frac{1}{2}$, $m_{a}=+\frac{1}{2}$, and $m_{b}=+\frac{1}{2}$, where $|\Delta|$ is the magnitude of the difference between equations (\ref{eq: final}) and (\ref{eq: Wigner}). In both panels $l_{a}$ is treated as a nuisance variable and the angle is assigned by the calibration (\ref{eq: theta}). Bottom: the maximum of $|\Delta|$ over $\theta_{ab}$ as a function of $n$, for the three systems treated in this paper; the dashed line is proportional to $n^{-1}$. \label{fig: spin 1/2 plot}}
\end{figure}

\begin{figure}
  \begin{centering}
    \includegraphics[scale=1]{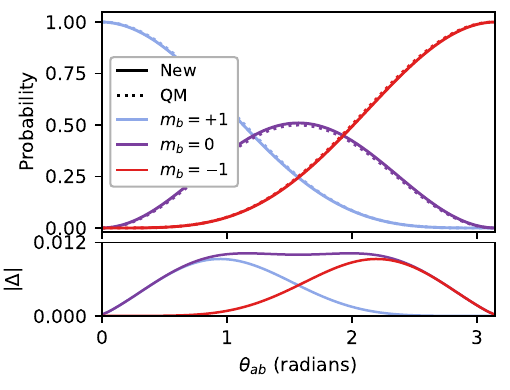}
    \hspace{1cm}
    \includegraphics[scale=1]{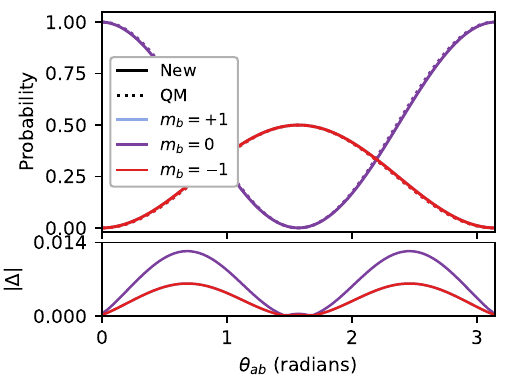}
  \end{centering}
  \caption{A comparison of models for $n=100$, $j=1$, $m_{a}=+1$ (top), and $m_{a}=0$ (bottom), where $|\Delta|$ is the magnitude of the difference between equations (\ref{eq: final}) and (\ref{eq: Wigner}) and $l_{a}$ is treated as a nuisance variable.\label{fig: j=1}}
\end{figure}

Given a system with total spin $j$, spin projection $m_{a}$ observed at Alice's detector, and a relative rotation of spatial frames $\theta_{ab}$ between Alice's and Bob's detectors, the probability that Bob observes a particular spin projection  $m_{b}$ in the proposed model is given by the following expression, where $P(r|u)\in \mathbb{Q}\cap \left[0,1\right]$ for finite $n$: 

\begin{equation}
    P(r|u)=
    \frac{\sum_{w\in Q(r,u)}\Upsilon_a(r,u,w) \Upsilon_b(r,u,w)}{\sum_{(r,w) \in Q(u)}\Upsilon_a(r,u,w) \Upsilon_b(r,u,w)}
    \label{eq: final}
\end{equation}

Using the same notation, the analogous expression from non-relativistic QM is as follows, where we have applied the Born rule to Wigner's d-matrix element:

\begin{equation}
\scalebox{.65}{$\begin{aligned}
&\left(d_{m_{b},m_{a}}^{j}(\theta_{ab})\right)^{2}= \\
&\sum_{q^{a}}(-1)^{m_{b}-m_{a}+q^{a}}\frac{(j+m_{a})!(j-m_{a})!}{(j+m_{a}-q^{a})!q^{a}!(m_{b}-m_{a}+q^{a})!(j-m_{b}-q^{a})!}\\
&\times\left(cos(\frac{\theta_{ab}}{2})\right)^{2j+m_{a}-m_{b}-2q^{a}}\left(sin(\frac{\theta_{ab}}{2})\right)^{m_{b}-m_{a}+2q^{a}}\\
&\times\sum_{q^{b}}(-1)^{m_{b}-m_{a}+q^{b}}\frac{(j+m_{b})!(j-m_{b})!}{(j+m_{a}-q^{b})!q^{b}!(m_{b}-m_{a}+q^{b})!(j-m_{b}-q^{b})!}\\
&\times\left(cos(\frac{\theta_{ab}}{2})\right)^{2j+m_{a}-m_{b}-2q^{b}}\left(sin(\frac{\theta_{ab}}{2})\right)^{m_{b}-m_{a}+2q^{b}}
\end{aligned}$}
\label{eq: Wigner}
\end{equation}

The expression in equation (\ref{eq: Wigner}) can be obtained in a variety of ways within the formalism of QM. It was first proposed by Eugene Wigner in 1927, who relied on group theoretic arguments \citep{Wigner1927, Wigner1932}. Decades later, Julian Schwinger provided an alternative derivation employing the operator algebra of simple harmonic oscillators in QM \citep{Schwinger1965}. A succinct outline of this derivation, which employs the Baker–Campbell–Hausdorff formula, can be found in \citep{Sakurai19941995}. A comparison of these two expressions for the cases of spin $\frac{1}{2}$ and $1$ is offered in Figures \ref{fig: spin 1/2 plot} and \ref{fig: j=1}, where a small deviation is observed. This can be made smaller by treating $l_{a}$ as a conditioning variable and using it to tune the the generated probability distribution, as seen in the bottom of Figure \ref{fig: spin 1/2 plot}. The difference can then be made arbitrarily small by increasing $n$ and adjusting $l_{a}$.

To be clear, the steps taken after equation (\ref{eq: base-4 cardinality}) were motivated by pragmatism. Simply put, they were the steps necessary to produce a result that closely matched the behavior of the modeled system. This leaves many open questions about the proposed model. These include the physical interpretation of the quantum numbers $n$ and $l$, the true nature of the ontic states being counted by the product $\Upsilon_a\Upsilon_b$, interference, and many others. Importantly, these questions arise in the context of a novel formalism that clearly has a deep connection to the modeled physical system, as can be seen in Figures \ref{fig: spin 1/2 plot} and \ref{fig: j=1}. It is also worth mentioning that a model for spin addition, which includes the rules for angular momentum addition and the Clebsch-Gordon coefficients, can be found by replacing the constraint $\tilde{C}_{map}=\tilde{D}_{map}=0$ with $\tilde{C}_{map}=\tilde{B}_{map}=0$ \citep{Powers2022, powers2023}.

For a full century, proposed interpretations of quantum systems have been bounded by the formal tools of canonical quantum theory, such as Hilbert spaces, the Born rule, and the Schr{\"o}dinger equation. In the author's view, the most exciting aspect of this work is the opportunity to develop a new picture of the modeled physical system. In the next section, we will begin the pedagogical introduction of this new model. Throughout the development process, the formal elements introduced in this section will be assigned a clear physical interpretation whenever possible. Of course, these interpretations should be viewed as provisional in these early stages of model development.

\section{Changing variables\label{sec:Changing-variables}}

We begin the pedagogical development of the formalism and model introduced in section \ref{sec: Formalism} by executing a change of variables in equation (\ref{eq: Wigner}), which is the result provided by QM. In Schwinger's oscillator model for this physical system, he begins with two uncoupled simple harmonic oscillators which may be called ``plus type'' and ``minus type'' \citep{Sakurai19941995}. Ladder operators associated with a particular spatial frame are then introduced for each oscillator, such that operators acting on different oscillators commute. The total angular momentum $j$ of a physical system is then built up from the vacuum state by successive applications of creation operators, which are denoted as $a_{+}^{\dagger}$ and $a_{-}^{\dagger}$
for the plus and minus type oscillator, respectively. Within the chosen spatial frame, the projection of angular momentum $m_{a}$ is then defined as the difference between the number of $a_{+}^{\dagger}$'s and $a_{-}^{\dagger}$'s used to lift the state from vacuum, where we continue to use the subscript notation introduced in section \ref{sec: Formalism}. Both $j$ and $m_{a}$ can be expressed in terms of the eigenvalue of the number operators for each oscillator type:

\begin{equation}
j\equiv\frac{n_{+}+n_{-}}{2}\label{eq: Schwinger j}
\end{equation}
\begin{equation}
m_{a}\equiv\frac{n_{+}-n_{-}}{2}\label{eq: Schwinger m}
\end{equation}

A second set of ladder operators can then be defined as rotated versions
of the originals. In this case, the $a_{+}^{\dagger}$
operator becomes $a_{+}^{\dagger}cos(\frac{\theta}{2})+a_{-}^{\dagger}sin(\frac{\theta}{2})$
and the $a_{-}^{\dagger}$ operator becomes $a_{-}^{\dagger}cos(\frac{\theta}{2})-a_{+}^{\dagger}sin(\frac{\theta}{2})$.
This implies that the projection of angular momentum in the rotated
spatial frame may be different than in the unrotated frame, which
can be defined in terms of the rotated number operator eigenvalues:

\begin{equation}
m_{b}\equiv\frac{n_{+}'-n_{-}'}{2}\label{eq: Schwinger m'}
\end{equation}

The expressions given in equations (\ref{eq: Schwinger j}-\ref{eq: Schwinger m'}) are structurally identical to those introduced in equation (\ref{eq: base-4 quantum numbers}). However, we no longer interpret the positive integers $n_{+},$ $n_{-}$, $n_{+}'$, and $n_{-}'$ as the eigenvalues of number operators for simple harmonic oscillators. Rather, we think of them as the number of times an abstract symbol appears within a sequence, suppressing any presuppositions about their physical meaning. This interpretation is motivated by the combinatorial terms in equation (\ref{eq: Wigner}), which are restated here:

\begin{equation}
\scalebox{.9}{$\begin{aligned}
\frac{(j+m_{a})!(j-m_{a})!}{(j+m_{a}-q^a)!q^a!(m_{b}-m_{a}+q^a)!(j-m_{b}-q^a)!}
\end{aligned}$}
\label{eq: alice base-4 counts}
\end{equation}

\begin{equation}
\scalebox{.9}{$\begin{aligned}
\frac{(j+m_{b})!(j-m_{b})!}{(j+m_{a}-q^b)!q^b!(m_{b}-m_{a}+q^b)!(j-m_{b}-q^b)!}
\end{aligned}$}
\label{eq: bob base-4 count}
\end{equation}

As we will show, each of these expressions can be interpreted as the cardinality of a set of base-4 sequences. We denote these sets as $\varepsilon^{CD}_{a}$ for equation (\ref{eq: alice base-4 counts}) and $\varepsilon^{CD}_{b}$ for equation (\ref{eq: bob base-4 count}). These sets are closely related to the contextual sets introduced in equation (\ref{eq: contextual sets}), but only include sequences containing the symbols $CC$, $CD$, $DC$, and $DD$, as indicated by the superscripts. A simple proof that this interpretation is valid can be found by summing the arguments in the numerator and denominator, separately: 

\begin{equation}
    (j + m_{a}) + (j - m_{a})=2j
\end{equation}
\begin{equation}
    \begin{aligned}
        (j+m_{a}-q)+(q)+(m_{b}-m_{a}+q)&\\
        +(j-m_{b}-q)&=2j
    \end{aligned}
\end{equation}

These sums both equal $2j\in \mathbb{N}_0$, which we may interpret as the length of the base-4 sequences comprising $\varepsilon^{CD}_a$ and $\varepsilon^{CD}_b$.  This length would typically be the only term in the numerator, as with the base-2 sequence counting binomial coefficient:

\begin{equation}
    \frac{n!}{k!(n-k)!}
\end{equation}

The presence of two arguments in the numerator of equation (\ref{eq: alice base-4 counts}) and (\ref{eq: bob base-4 count}) implies a constraint on the elements of $\varepsilon^{CD}_a$ and $\varepsilon^{CD}_b$. This constraint motivates us to interpret each base-4 sequence as a Cartesian product ($\otimes$) of two base-2 sequences, each comprised of the symbols $C$ and $D$. An example of one such construction is given below, where the subscripts $a1$ and $b2$ distinguish the left and right base-2 sequence, respectively:

\begin{equation}
\scalebox{0.8}{$\begin{aligned}
\left(\begin{array}{c}
C\\
D\\
D\\
C
\end{array}\right)_{a}\otimes\left(\begin{array}{c}
C\\
D\\
C\\
D
\end{array}\right)_{b}=\left(\begin{array}{c}
CC\\
DD\\
DC\\
CD
\end{array}\right)_{a1.b2}
\end{aligned}$}
\label{eq: base-4 example}
\end{equation}

The constraint imposed by the two factorials in the numerator of equation
(\ref{eq: alice base-4 counts}) and (\ref{eq: bob base-4 count})
can now be interpreted as holding one of the base-2 sequences fixed
while counting base-4 permutations. That is, all base-4 sequences in the set $\varepsilon^{CD}_a$ must share a common $a1$ base-2 sequence. Likewise, all base-4 sequences in the set $\varepsilon^{CD}_b$ must share a common $b2$ base-2 sequence. To make this picture more clear, we will show that all the arguments in the combinatorial expressions (\ref{eq: alice base-4 counts}) and (\ref{eq: bob base-4 count}) can be interpreted as the number of times a symbol appears in a base-4 sequence. We call these numbers counts and denote them by placing a tilde atop the symbol of interest like so: 

\begin{equation}
    \widetilde{CC}, \widetilde{CD},\widetilde{DC},\widetilde{DD}\in \mathbb{N}_0 
\end{equation}

 Following from the construction
depicted in equation (\ref{eq: base-4 example}), we can express the
base-2 counts associated with the $a1$ and $b2$ sequences in terms
of base-4 counts like so:

\begin{equation}
\tilde{C}_{a}=\widetilde{CC}+\widetilde{CD},\qquad\tilde{D}_{a}=\widetilde{DC}+\widetilde{DD}\label{eq: C_a1 D_a1}
\end{equation}
\begin{equation}
\tilde{C}_{b}=\widetilde{CC}+\widetilde{DC},\qquad\tilde{D}_{b}=\widetilde{CD}+\widetilde{DD}\label{eq: C_b2 D_b2}
\end{equation}

The positive integers $n_{+}$, $n_{-}$, $n_{+}'$,
and $n_{-}'$ interpreted by Schwinger as number operator eigenvalues
will now be interpreted as base-2 counts:

\begin{equation}
n_{+}\rightarrow\tilde{C}_{a},\qquad n_{-}\rightarrow\tilde{D}_{a}\label{eq: n+-}
\end{equation}
\begin{equation}
n'_{+}\rightarrow\tilde{C}_{b},\qquad n'_{-}\rightarrow\tilde{D}_{b}\label{eq: n+-'}
\end{equation}

This change maps the definitions of $j$, $m_{a}$, and $m_{b}$ given in equations (\ref{eq: Schwinger j}-\ref{eq: Schwinger m'}) to the definitions given in equation (\ref{eq: base-4 quantum numbers}). One last definition is necessary to accomplish our goal of interpreting each factorial argument as a count. This definition is for the summing parameter $q$, which will carry the superscript $a$ or $b$ when calculating the cardinality of  $\varepsilon^{CD}_a$ or $\varepsilon^{CD}_b$, respectively:

\begin{equation}
q^{a}\equiv\widetilde{CD}^{a},\qquad q^{b}\equiv\widetilde{CD}^{b}\label{eq: q}
\end{equation}

Using the definitions in equations (\ref{eq: base-4 quantum numbers}), (\ref{eq: C_a1 D_a1}), (\ref{eq: C_b2 D_b2}), and (\ref{eq: q}), we may now execute the following change of variables for the remaining factorial arguments in equations (\ref{eq: alice base-4 counts}) and (\ref{eq: bob base-4 count}):

\begin{equation}
j+m_{a}=\tilde{C}_{a},\qquad j-m_{a}=\tilde{D}_{a}\label{eq: Alice num}
\end{equation}
\begin{equation}
j+m_{b}=\tilde{C}_{b},\qquad j-m_{b}=\tilde{D}_{b}\label{eq: Bob num}
\end{equation}

\begin{equation}
\begin{aligned}
    j+m_{a}-q^{a}=\widetilde{CC}^{a}\\
    m_{b}-m_{a}+q^{a}=\widetilde{DC}^{a}\\
    j-m_{b}-q^{a}=\widetilde{DD}^{a}
\end{aligned} 
    \label{eq: Alice den}  
\end{equation}

\begin{equation}
\begin{aligned}
    j+m_{a}-q^{b}=\widetilde{CC}^{b}\\
    m_{b}-m_{a}+q^{b}=\widetilde{DC}^{b}\\
    j-m_{b}-q^{b}=\widetilde{DD}^{b}
\end{aligned} 
    \label{eq: Bob den} 
\end{equation}

With these results, equations (\ref{eq: alice base-4 counts}) and
(\ref{eq: bob base-4 count}) become the following:

\begin{equation}
\frac{\tilde{C}_{a}!\tilde{D}_{a}!}{\widetilde{CC}^{a}!\widetilde{CD}^{a}!\widetilde{DC}^{a}!\widetilde{DD}^{a}!}\label{eq: alice base-4 counts new}
\end{equation}
\begin{equation}
\frac{\tilde{C}_{b}!\tilde{D}_{b}!}{\widetilde{CC}^{b}!\widetilde{CD}^{b}!\widetilde{DC}^{b}!\widetilde{DD}^{b}!}\label{eq: bob base-4 count new}
\end{equation}

Expressed in this form, the proposed interpretation of these combinatorial
terms becomes more clear. We may think of them as counting base-4
sequences, such that the ordering of the symbols in one of the component
base-2 sequences remain fixed. In the case of equation (\ref{eq: alice base-4 counts new}),
it is the base-2 sequence with the subscript $a1$ which is held fixed,
while the base-2 sequence with the subscript $b2$ is held fixed in
equation (\ref{eq: bob base-4 count new}). One example of the elements in $\varepsilon^{CD}_a$ being counted by equation (\ref{eq: alice base-4 counts new}) is as follows, where $\tilde{C}_{a}=\tilde{D}_{a}=\tilde{C}_{b}=\tilde{D}_{b}=2$ and $\widetilde{CC}^{a}=\widetilde{CD}^{a}=\widetilde{DC}^{a}=\widetilde{DD}^{a}=1$:

\begin{equation}
\scalebox{0.8}{$\begin{aligned}
\left(\begin{array}{c}
C\\
D\\
D\\
C
\end{array}\right)_{a}\otimes\left\{ \left(\begin{array}{c}
C\\
D\\
C\\
D
\end{array}\right)_{b},\left(\begin{array}{c}
D\\
D\\
C\\
C
\end{array}\right)_{b},\left(\begin{array}{c}
C\\
C\\
D\\
D
\end{array}\right)_{b},\left(\begin{array}{c}
D\\
C\\
D\\
C
\end{array}\right)_{b}\right\}
\end{aligned}$}
\label{eq: Alice's ensemble}
\end{equation}

For the example given above, the combinatorial expression in equation
(\ref{eq: alice base-4 counts new}) yields:

\[
\frac{\tilde{C}_{a}!\tilde{D}_{a}!}{\widetilde{CC}^{a}!\widetilde{CD}^{a}!\widetilde{DC}^{a}!\widetilde{DD}^{a}!}=\frac{2!2!}{1!1!1!1!}=4
\]

Motivated by this combinatorial picture, we propose that the sets $\varepsilon^{CD}_a$ and $\varepsilon^{CD}_b$ be interpreted as statistical ensembles associated with observers named Alice (a) and Bob (b). The base-4 sequences comprising $\varepsilon^{CD}_a$ and $\varepsilon^{CD}_b$ are then interpreted as the possible ontic states of the physical system being modeled. As depicted in Figure \ref{fig: sketch 1}, Alice and Bob are each assigned to one of the SG detectors involved in the experiment of interest. Through the use of their detector, each observer has access to one component (base-2 sequence) of the full ontic state (base-4 sequence). We interpret these components as the measurement events that occur in each observer's detector, which we distinguish from one another with the subscripts $a1$ (Alice) and $b2$ (Bob). An observer's statistical ensemble is then constructed by holding their own measurement event fixed while considering all possible measurement events that may occur at the other observer's detector. Under this interpretation, the hidden information that gives rise to non-determinism is stored in the ordering of the base-4 symbols in each ontic state. \textit{That is, counts are generally observable, but sequences are not.} 

A clear shortcoming of the proposed interpretation is the absence of the quantity $\theta$ in the combinatorial terms we have been studying. The sine and cosine terms in equation (\ref{eq: Wigner}) not only account for rotation, but they also serve to normalize the combinitorial expressions. In the coming sections, we will replace these terms by adding the symbols $A$ and $B$ to the alphabet used to model measurement events. The result will be an expanded version of equations (\ref{eq: alice base-4 counts new}) and (\ref{eq: bob base-4 count new}), which can be normalized in the fashion of a relative frequency. The inclusion of these new symbols also allows measurement events to be expressed as Cartesian products of base-2 sequences composed of the symbols $0$ and $1$. This is the origin of the $a1$ and $b2$ subscript notation employed in this section and is central to the issues of locality and realism. Specifically, the decomposition of measurement events into two components highlights the relational nature of the quantum numbers $m_{a}$ and $m_{b}$.

Two important aspects of equation (\ref{eq: Wigner}) that we have not yet discussed are the interference terms and the Born rule. First, we note that the summing parameters $q^a=\widetilde{CD}^a$ and $q^b=\widetilde{CD}^b$ are responsible for driving interference. Unlike $j$, $m_{a}$, and $m_{b}$, these quantities are not observable properties of events. Rather, they are properties of the relationship between events. Though we lack the development to fully illustrate this point now, it will be addressed directly in section \ref{sec: Spin 1 arbitrary n}. As interference phenomena are intimately related to path lengths through spacetime, the new perspective afforded by the proposed model is of broad interest.

Finally, we point out that the Born rule, which was employed in the construction of equation (\ref{eq: Wigner}), is directly responsible for the presence of both $\varepsilon^{CD}_a$ and $\varepsilon^{CD}_b$. Within the proposed model, it suggests that the full ontic state space of interest when calculating probabilities is actually the Cartesian product of Alice's and Bob's statistical ensembles. Simply put, the Born rule tells us to count the number of unique ways to move between Alice's and Bob's statistical ensembles. In the coming sections, the full picture associated with these statistical ensembles and their Cartesian product will be developed.

\section{Spin 1/2 $\quad \theta \in \left\{0,\pi \right\}$ \label{sec: Spin 1/2 n=1}}

\begin{figure}
\begin{centering}
\includegraphics[scale=.65]{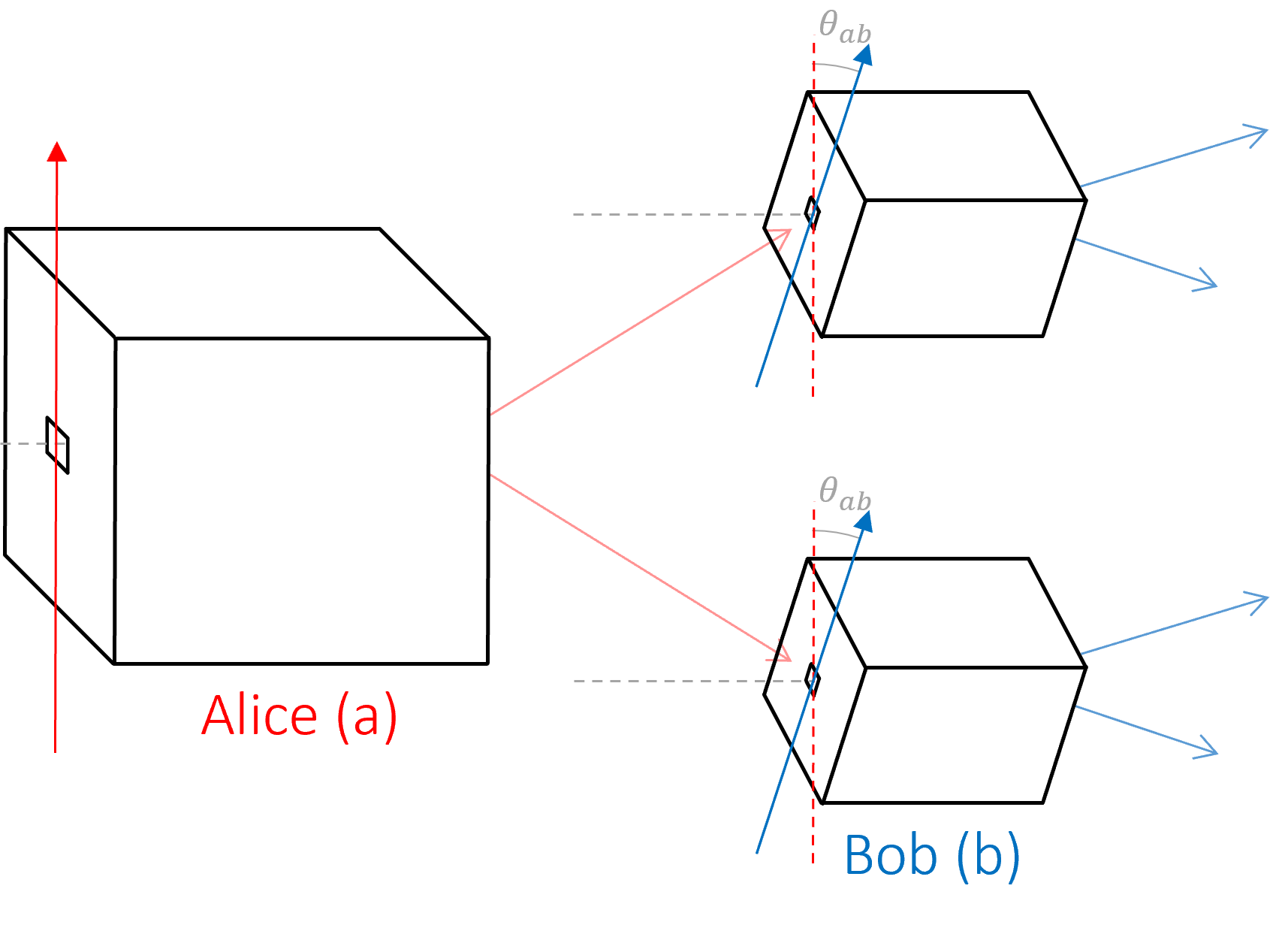}
\par\end{centering}
\caption{A measurement event occurs at Alice's detector which deflects the $j=\frac{1}{2}$ particle into one of two paths (red), one for each possible value of the quantum number $m_{a}\in \{+\frac{1}{2},-\frac{1}{2}\}$. Depending on which value of $m_{a}$ is of interest, Bob's detector is then placed along one of these paths. Bob then rotates his detector with respect to Alice's by the angle $\theta_{ab}$. Finally, a measurement event occurs at Bob's detector which again deflects the $j=\frac{1}{2}$ particle into one of two paths (blue), one for each possible value of the quantum number $m_{b}\in \{+\frac{1}{2},-\frac{1}{2}\}$.}\label{fig: sketch 2}
\end{figure}

In the previous section, we made the suggestion that sequences comprised of the abstract symbols $C$ and $D$ should be interpreted as models for measurement events. To clarify what we mean by this, we will treat the simplest possible case of an experiment involving two SG detectors, which is for a spin $\frac{1}{2}$ particle. In Figure \ref{fig: sketch 2}, we see an example of such an experiment. The first event that takes place in this experiment is the interaction, or measurement event at Alice's detector. The result of this measurement event is the deflection of the spin $\frac{1}{2}$ particle into one of two possible paths, one for each value of $m_{a}\in \{+\frac{1}{2} , -\frac{1}{2}\}$. For the spin $\frac{1}{2}$ case, the only base-2 sequences that could be associated with this event, each of which have length $n=1$, are as follows:

\begin{equation}
\left\{ \left(\begin{array}{c}
C\end{array}\right)_{a},\:\left(\begin{array}{c}
D\end{array}\right)_{a}\right\}
\label{eq: Alice spin 1/2 n=1}
\end{equation}

Collectively, these sequences form the ontic state space for Alice's measurement event. That is, we assume that the event in Alice's detector is associated with a definite state of reality, which we model using a single sequence of abstract symbols. Continuing to follow the spin $\frac{1}{2}$ particle's path in Figure \ref{fig: sketch 2}, we see that there will also be a measurement event in Bob's detector. Like before, this event will cause the particle to be deflected into one of two possible paths. As we did with Alice's event, we can write down the ontic state space for Bob's event like so:

\begin{equation}
\left\{\left(\begin{array}{c}
C\end{array}\right)_{b},\:\left(\begin{array}{c}
D\end{array}\right)_{b}\right\}
\label{eq: Bob spin 1/2 n=1}
\end{equation}

For any given experiment, an event occurs at Alice's and Bob's detector. Thus, the ontic state space for the modeled system should consist of all ordered pairs of sequences in the ontic state spaces for each measurement event. That's just the Cartesian product of equations (\ref{eq: Alice spin 1/2 n=1}) and (\ref{eq: Bob spin 1/2 n=1}):

\begin{equation}
\left(\begin{array}{c}
C\end{array}\right)_{a}\otimes\left(\begin{array}{c}
C\end{array}\right)_{b}=\left(\begin{array}{c}
CC\end{array}\right)_{a,b}
\label{eq: spin 1/2 n=1 example-1}
\end{equation}
\begin{equation}
\left(\begin{array}{c}
C\end{array}\right)_{a}\otimes\left(\begin{array}{c}
D\end{array}\right)_{b}=\left(\begin{array}{c}
CD\end{array}\right)_{a,b}\label{eq: spin 1/2 n=1 example-2}
\end{equation}
\begin{equation}
\left(\begin{array}{c}
D\end{array}\right)_{a}\otimes\left(\begin{array}{c}
C\end{array}\right)_{b}=\left(\begin{array}{c}
DC\end{array}\right)_{a,b}\label{eq: spin 1/2 n=1 example-3}
\end{equation}
\begin{equation}
\left(\begin{array}{c}
D\end{array}\right)_{a}\otimes\left(\begin{array}{c}
D\end{array}\right)_{b}=\left(\begin{array}{c}
DD\end{array}\right)_{a,b}\label{eq: spin 1/2 n=1 example-4}
\end{equation}

We are now ready to begin discussing rotation, which we may think of as an operation which maps Alice's event to Bob's. The extreme cases of rotation are $\theta_{ab}=0$ and $\theta_{ab}=\pi$, which should correspond to $m_{a}=m_{b}$ and $m_{a}=-m_{b}$, respectively. Returning to the ontic state space of the experiment given in equations (\ref{eq: spin 1/2 n=1 example-1}-\ref{eq: spin 1/2 n=1 example-4}), we see that equations (\ref{eq: spin 1/2 n=1 example-1}) and (\ref{eq: spin 1/2 n=1 example-4}) are associated with the $\theta_{ab}=0$ case ($m_{a}=m_{b}$), while equations (\ref{eq: spin 1/2 n=1 example-2}) and (\ref{eq: spin 1/2 n=1 example-3}) are associated with the $\theta_{ab}=\pi$ case ($m_{a}=-m_{b}$).

Because we want to think of rotation as an operation, we need to assign the symbols $C$ and $D$ some algebraic property which allows them to be mapped into one another. There are many ways to approach this, but in the end, making the symbols $C$ and $D$ elements of the finite Klein four-group ($Z_{2}\times Z_{2}$) will prove most fruitful. Because there are four elements in this group, we need to add the symbols $A$ and $B$ to our alphabet for modeling events. What remains is to decide which elements of this group to associate with each of the four symbols. These assignments, which were given in equation (\ref{eq: base-4 definitions}), are restated here:

\begin{equation}
    \begin{aligned}
        A_{a}\equiv 0_a\otimes 0_1, \quad
        B_{a}\equiv 1_a\otimes 1_1\\
        C_{a}\equiv 1_a\otimes 0_1, \quad
        D_{a}\equiv 0_a\otimes 1_1
    \end{aligned}
\end{equation}

Note that an important consequence of using the finite group $Z_{2}\times Z_{2}$ is that we may actually regard base-4 sequences composed of the symbols $A$, $B$, $C$, and $D$ as Cartesian products ($\otimes$) of base-2 sequences composed of the symbols $0$ and $1$. This is the origin of the subscript notation we have been using thus far, where the indices $a$ and $1$, for example, distinguish the base-2 sequences within the ordered pair. An example of this construction is given here:

\begin{equation}
\scalebox{0.8}{$\begin{aligned}
  \left(\begin{array}{c}
0\\
1\\
1\\
0\\
1\\
0
\end{array}\right)_{a}\otimes\left(\begin{array}{c}
0\\
1\\
0\\
1\\
1\\
0
\end{array}\right)_{1}=\left(\begin{array}{c}
A\\
B\\
C\\
D\\
B\\
A
\end{array}\right)_{a}
\end{aligned}$}
\label{eq: base-2 correlation}
\end{equation}

The group operation of interest for rotations is element-wise addition modulo two, which we denote using the symbol $\oplus$. In computer science, this operation is referred to as $XOR$. In general, the operators which map ontic states for Alice's measurement event into Bob's are sequences comprised of the symbols $A$, $B$, $C$, and $D$, along with the $\oplus$ operation. Though, for rotations, we are only concerned with sequences filled with $A$'s and $B$'s. This restriction is motivated in part by the need for Alice's and Bob's events to share a common value of $j$, a restriction that may be lifted for systems other than the one of interest here. The full addition table for these symbols is as follows:

\begin{equation}
\begin{array}{c}
A\oplus A=B\oplus B=C\oplus C=D\oplus D=A\\
A\oplus B=B\oplus A=C\oplus D=D\oplus C=B\\
A\oplus C=C\oplus A=B\oplus D=D\oplus B=C\\
A\oplus D=D\oplus A=B\oplus C=C\oplus B=D
\end{array}\label{eq: base-4 addition table}
\end{equation}

With our brief interlude into finite group theory complete, we can return to the physical system we are attempting to model. The goal now is to show that the ontic state space for the experiment, which was defined in equations (\ref{eq: spin 1/2 n=1 example-1}-\ref{eq: spin 1/2 n=1 example-4}), can be generated by applying maps to the ontic state space of Alice's measurement event, for example. For the case we have been considering, there are two ontic states associated with Alice's event and two possible maps for each, yielding four possible scenarios: 

\begin{equation}
\left(\begin{array}{c}
C\end{array}\right)_{a}\oplus\left(\begin{array}{c}
A\end{array}\right)_{map}=\left(\begin{array}{c}
C\end{array}\right)_{b}\label{eq: base-4 map 1}
\end{equation}
\begin{equation}
\left(\begin{array}{c}
C\end{array}\right)_{a}\oplus\left(\begin{array}{c}
B\end{array}\right)_{map}=\left(\begin{array}{c}
D\end{array}\right)_{b}\label{eq: base-4 map 2}
\end{equation}
\begin{equation}
\left(\begin{array}{c}
D\end{array}\right)_{a}\oplus\left(\begin{array}{c}
B\end{array}\right)_{map}=\left(\begin{array}{c}
C\end{array}\right)_{b}\label{eq: base-4 map 3}
\end{equation}
\begin{equation}
\left(\begin{array}{c}
D\end{array}\right)_{a}\oplus\left(\begin{array}{c}
A\end{array}\right)_{map}=\left(\begin{array}{c}
D\end{array}\right)_{b}\label{eq: base-4 map 4}
\end{equation}

Now that we have an understanding of how maps work, we need to make a choice regarding the definition of $\theta_{ab}$ in terms of these maps. Taking a look at equations  (\ref{eq: base-4 map 1}-\ref{eq: base-4 map 4}), we see that the $m_{a}=m_{b}$  case is associated with the symbol $A$ ($\theta_{ab}=0$), while the $m_{a}=-m_{b}$  case is associated with the symbol $B$ ($\theta_{ab}=\pi$). Any acceptable rule relating $\theta_{ab}$ to the map counts must therefore return $\theta_{ab}=0$ when $\tilde{B}_{map}=0$ and $\theta_{ab}=\pi$ when $\tilde{A}_{map}=0$, where $\tilde{B}_{map}$ is the number of $B$'s that appear in a map of length $n$ and $\tilde{A}_{map}=n-\tilde{B}_{map}$. These two extreme cases do not by themselves determine the rule at intermediate angles. The correct assignment is the half-angle law

\begin{equation}
\tan\left(\frac{\theta_{ab}}{2}\right)\equiv\frac{\tilde{B}_{map}}{\tilde{A}_{map}}\label{eq: theta}
\end{equation}

Two independent arguments fix this form, and neither leaves a free parameter. The first is structural. The rotation maps form a one parameter family labelled by the pair of counts $(\tilde{A}_{map},\tilde{B}_{map})$ subject to $\tilde{A}_{map}+\tilde{B}_{map}=n$, and any acceptable angle assignment must be compatible with the composition of successive rotations: if a map taking Alice's frame to Bob's is followed by a map taking Bob's frame to a third frame, the assigned angles must add. Associating to each map the complex number $z\equiv\tilde{A}_{map}+i\tilde{B}_{map}$, and to the composition of two maps the product $z_1z_2$ rescaled so that the composite again has counts summing to $n$, the arguments add exactly. The only angle-like quantity carried by a map which is additive under composition is therefore $\arg(z)=\arctan(\tilde{B}_{map}/\tilde{A}_{map})$. A rule linear in $\tilde{B}_{map}$ is not additive under this composition and cannot be made so by any reparametrisation. What remains is to fix the proportionality between $\arg(z)$ and the physical angle, and this is settled by the fact that the objects being rotated are spinors: as seen in section \ref{sec:Changing-variables}, the rotation acts on the oscillator algebra through the half angle $\theta_{ab}/2$. Identifying the additive map angle with the spinor half angle gives equation (\ref{eq: theta}).

The second argument is a direct calculation, and we return to it once the general expression for probabilities has been assembled. For $j=\frac{1}{2}$ and $m_{a}=+\frac{1}{2}$ the block of counts carrying the symbols $C$ and $D$ contains a single unit, so each contextual set contains exactly one ontic state and interference is absent. In that case equation (\ref{eq: final}) can be evaluated in closed form, and taking $n\rightarrow\infty$ at fixed $\tilde{B}_{map}/n$ yields

\begin{equation}
\lim_{n\rightarrow\infty}P\left(m_{b}=-\tfrac{1}{2}\right)=\frac{\tilde{B}_{map}^2}{\tilde{A}_{map}^2+\tilde{B}_{map}^2}
\label{eq: half limit}
\end{equation}

The corresponding prediction of QM is $\sin^2(\theta_{ab}/2)$, and equating the two requires precisely equation (\ref{eq: theta}). We are careful not to overstate the first argument: the composition rule assumed there is an additional structure imposed on the map counts, and is not derived from the $\oplus$ operation, since the counts of a composition of two maps under $\oplus$ depend on the overlap of the two sequences and not on their counts alone. What it establishes is that \emph{if} map angles are to be additive under composition, the assignment must take the form of equation (\ref{eq: theta}). That the same rule is independently forced by the spin $\frac{1}{2}$ sector is what settles the matter. Note also that equation (\ref{eq: theta}) gives $\cos\theta_{ab}=(\tilde{A}_{map}^2-\tilde{B}_{map}^2)/(\tilde{A}_{map}^2+\tilde{B}_{map}^2)$, so that $(\tilde{A}_{map},\tilde{B}_{map})$ generate the Pythagorean parametrisation of the unit circle.

We have finally developed enough machinery to check if the proposed model indeed predicts the correct behavior of a spin $\frac{1}{2}$ particle within an experiment involving two SG detectors. Though, we are currently limited to relative rotations of $\theta_{ab}=0$ and $\theta_{ab}=\pi$ between Alice's and Bob's detectors. In the case of $\theta_{ab}=0$, if Alice observes $m_{a}=+\frac{1}{2}$ ($m_{a}=-\frac{1}{2}$) then Bob must observe $m_{b}=+\frac{1}{2}$ ($m_{b}=-\frac{1}{2}$), as seen in equations (\ref{eq: base-4 map 1}) and (\ref{eq: base-4 map 4}). Likewise for $\theta_{ab}=\pi$, if Alice observes $m_{a}=+\frac{1}{2}$ ($m_{a}=-\frac{1}{2}$) then Bob must observe $m_{b}=-\frac{1}{2}$ ($m_{b}=+\frac{1}{2}$), as seen in equations (\ref{eq: base-4 map 2}) and (\ref{eq: base-4 map 3}). Of course, these results are unsurprising given the approach we took when selecting a definition for $\theta_{ab}$. The issue we must now face is that of arbitrary rotations.

\section{Spin 1/2   $\quad \theta \in \left\{0,\frac{\pi}{2}, \pi \right\}$ \label{sec: Spin 1/2 n=2}}

Continuing our incremental approach to developing this new model, we now consider the case in which Alice's and Bob's detector have a relative rotation of $\theta_{ab}=\frac{\pi}{2}$. From our definition of $\theta_{ab}$ given in equation (\ref{eq: theta}), a rotation of $\frac{\pi}{2}$ requires $\tilde{A}_{map}=\tilde{B}_{map}$, so the smallest value of $n$ that would support such a rotation is $2$. Given that we are modeling a spin $\frac{1}{2}$ particle, along with the definition of $j$ given in equation (\ref{eq: base-4 quantum numbers}), the total number of $C$'s and $D$'s that can appear in ontic states for measurement events is still limited to one. This leaves us with only two options for lengthening our sequences, which is adding an $A$ or a $B$. This modification results in the following ontic state spaces for Alice's and Bob's measurement event:

\begin{equation}
\scalebox{0.8}{$
\begin{aligned}
&\left\{ \left(\begin{array}{c}
A\\
C
\end{array}\right)_{a},\left(\begin{array}{c}
C\\
A
\end{array}\right)_{a},\left(\begin{array}{c}
B\\
C
\end{array}\right)_{a},\left(\begin{array}{c}
C\\
B
\end{array}\right)_{a},\right. \\
&\left. \left(\begin{array}{c}
A\\
D
\end{array}\right)_{a},\left(\begin{array}{c}
D\\
A
\end{array}\right)_{a},\left(\begin{array}{c}
B\\
D
\end{array}\right)_{a},\left(\begin{array}{c}
D\\
B
\end{array}\right)_{a}\right\}
\end{aligned}
$}
\label{eq: Alices n=2 state space}
\end{equation}

\begin{equation}
\scalebox{0.8}{$\begin{aligned}
&\left\{ \left(\begin{array}{c}
A\\
C
\end{array}\right)_{b},\left(\begin{array}{c}
C\\
A
\end{array}\right)_{b},\left(\begin{array}{c}
B\\
C
\end{array}\right)_{b},\left(\begin{array}{c}
C\\
B
\end{array}\right)_{b},\right. \\ 
&\left. \left(\begin{array}{c}
A\\
D
\end{array}\right)_{b},\left(\begin{array}{c}
D\\
A
\end{array}\right)_{b},\left(\begin{array}{c}
B\\
D
\end{array}\right)_{b},\left(\begin{array}{c}
D\\
B
\end{array}\right)_{b}\right\} 
\
\end{aligned}$}
\label{eq: Bobs n=2 state space}
\end{equation}

The inclusion of the symbols $A$ and $B$ in ontic states for measurement events has several important implications, but the first issue that must be addressed is the need to define two new quantum numbers, which we may think of as analogues of $j$ and $m$ (other definitions are possible):

\begin{equation}
g_{a}\equiv\frac{\tilde{A}_{a}+\tilde{B}_{a}}{2}\,,\qquad g_{b}\equiv\frac{\tilde{A}_{b}+\tilde{B}_{b}}{2}\label{eq: g}
\end{equation}
\begin{equation}
l_{a}\equiv\frac{\tilde{A}_{a}-\tilde{B}_{a}}{2}\,,\qquad l_{b}\equiv\frac{\tilde{A}_{b}-\tilde{B}_{b}}{2}\label{eq: l}
\end{equation}

We now have four quantum numbers associated with events, which are $j$, $m$, $g$, and $l$. However, we will often replace $g$ with $n$ when listing the degrees of freedom for events, where $n=2j+2g$. For the time being, we remain agnostic with respect to the physical interpretation of these new quantum numbers. When calculating the probability distribution for $m_{b}$, we will typically sum over all possible values of $l$. Though, as illustrated in Figure \ref{fig: spin 1/2 plot}, choosing particular values of $l_{a}$ has an interesting effect on the predictions generated by the new model. 

Beyond the new quantum numbers, the inclusion of $A$'s and $B$'s also lifts the one-to-one correspondence between quantum states and ontic states for spin $\frac{1}{2}$ particles. For example, there are now two sequences (ontic states) in equation (\ref{eq: Alices n=2 state space}) associated with each possible combination of $m_{a}$ and $l_{a}$ (quantum states). This change from one-to-one to one-to-many leads directly to non-determinism within this model, where the relative frequency of ontic states determines the probability of observing a particular quantum state. In the coming sections, we will introduce combinatorial tools to aid in the calculation of these relative frequencies. 

A further consequence of including $A$'s and $B$'s is that measurement events are no longer modeled by base-2 sequences comprised of the symbols $C$ and $D$. Rather, measurement events will now be modeled by base-4 sequences comprised of the symbols $A$, $B$, $C$, and $D$. This also implies that the full system of interest, which is the product of two measurement events, will now be modeled by base-16 sequences comprised of all possible pairs of these base-4 symbols. While this increase in basis may give the impression of this formalism becoming more complicated, the updated structure actually affords a more intuitive picture of the system being modeled. 

The new picture begins with the component base-2 sequences comprising measurement events. As illustrated in equation (\ref{eq: base-2 correlation}), measurement events can be interpreted as products of two base-2 sequences. To aid in the development of this new picture, we reintroduce the following set theory notation for individual base-2 sequences:

\begin{equation}
    s^1_{a},s^1_{b},s^1_{1},s^1_{2} \in S^1(n)
    \label{eq: set of all base-2 sequences}
\end{equation}

In equation (\ref{eq: set of all base-2 sequences}), $S^1(n)$ is the set of all base-2 sequences of length $n$, while $s_a^1$, $s_b^1$, $s_1^1$, and $s_2^1$ are individual elements of that set. Continuing with this notation, we can construct the base-4 sequences used to model Alice's and Bob's measurement events like so, where the ordering of component sequences within a given product depends on convention and physical context:  

\begin{equation}
    s^1_{a}\otimes s^1_{1},s^1_{1}\otimes s^1_{a},s^1_{b}\otimes s^1_{2},s^1_{2}\otimes s^1_{b} \in S^2(n)
    \label{eq: set of all base-4 sequences}
\end{equation}

In base-4 notation, these products are denoted as $s^2_{a1}$, $s^2_{1a}$, $s^2_{b2}$, and $s^2_{2b}$, respectively. These are the only places in this paper where the numeral index is displayed explicitly, since here it is the ordering itself that is at issue. Finally, we can construct products of measurement events, or base-16 sequences like so:

\begin{equation}
    s^2_{a}\otimes s^2_{b},s^2_{b}\otimes s^2_{a} \in S^4(n)
    \label{eq: set of all base-16 sequences}
\end{equation}

For any given measurement event, one of the component base-2 sequences can be associated with the observer of that event, which we denote using the subscript $a$ or $b$. We refer to these base-2 sequences as reference sequences and interpret them as models for the magnetic field gradients within Alice's and Bob's detectors, as depicted in Figure \ref{fig: sketch 2}. To model the physical process of a particle being deflected by Alice's magnetic field gradient, for example, we introduce the following base-2 sequence addition operation:

\begin{equation}
s^1_a\oplus s^1_1=s^1_j \qquad s^1_j \in S^1(n) 
\label{eq: alices measurement event operation}
\end{equation}

\begin{figure}
\begin{centering}
\includegraphics[scale=.4]{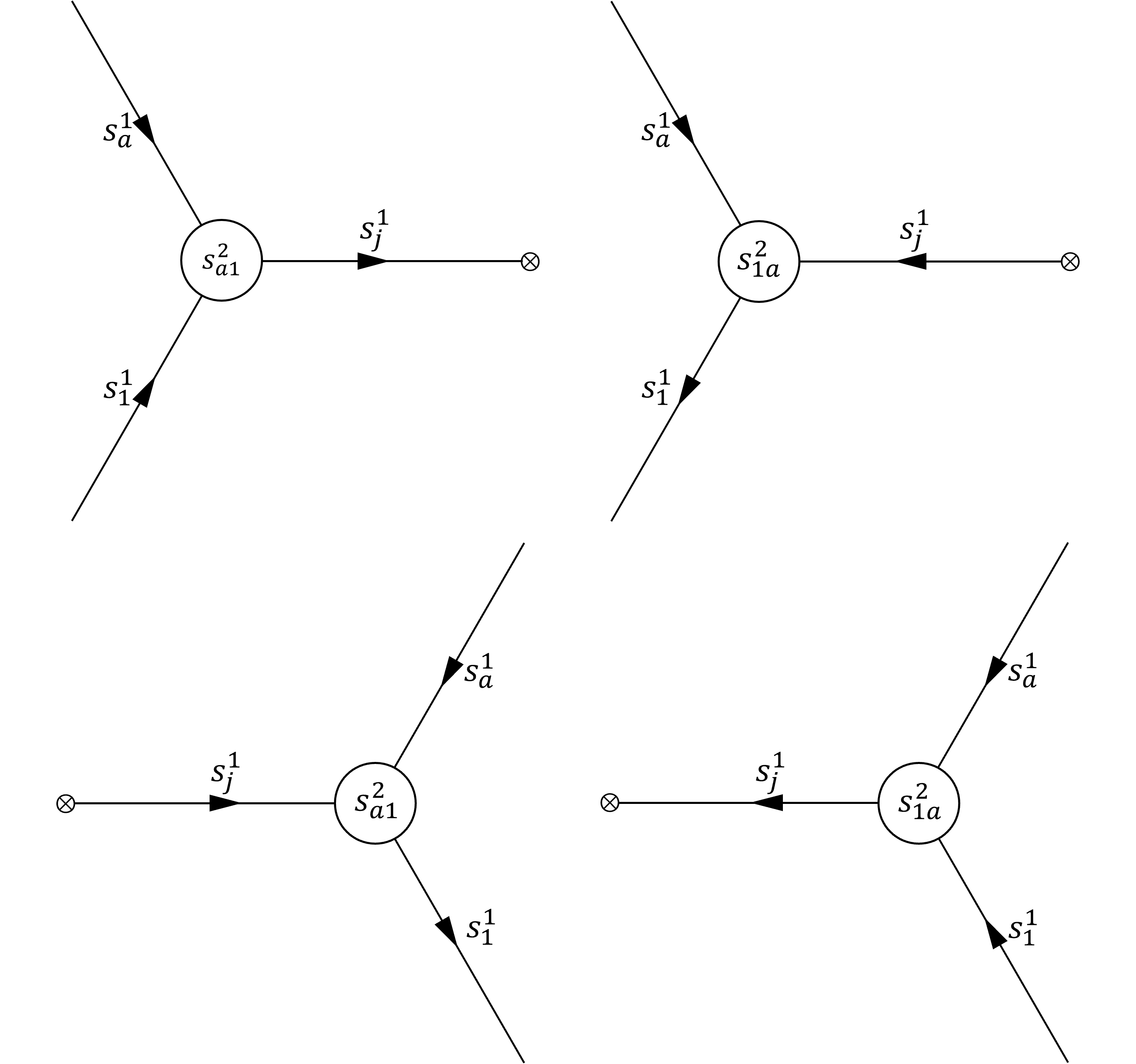}
\par\end{centering}
\caption{All possible graphical representations of Alice's measurement event. The ordering of the base-2 sequences $s^1_a$ and $s^1_1$ within Alice's measurement event determines the direction of the $s^1_j$ edge. The ``$a1$" ordering indicates a left to right direction for the $s^1_j$ edge, while the ``$1a$" ordering indicates a right to left direction. The position of the full base-4 sequences $s^2_{a1}$ or $s^2_{1a}$ with respect to the Cartesian product operator $\otimes$ distinguishes the edge configuration in the top row with the one in the bottom. Note that the numeral index is retained in this figure and in the accompanying discussion, where the ordering carries information; elsewhere it is suppressed, as described in section \ref{sec: Formalism}.}\label{fig: all of alices combinations Sketch}
\end{figure}

In equation (\ref{eq: alices measurement event operation}), the measurement event in Alice's detector is modeled by an addition operation involving Alice's reference sequence and a base-2 sequence carrying the subscript $1$. The result of this addition operation is a third base-2 sequence carrying the subscript $j$. We interpret this base-2 sequence as a model for the particle that connects Alice's and Bob's measurement events. A model for the interaction occurring in Bob's detector is given here:

\begin{equation}
s^1_b\oplus s^1_j=s^1_2
\label{eq: bobs measurement event operation}
\end{equation}

\begin{figure}
\begin{centering}
\includegraphics[width=\columnwidth]{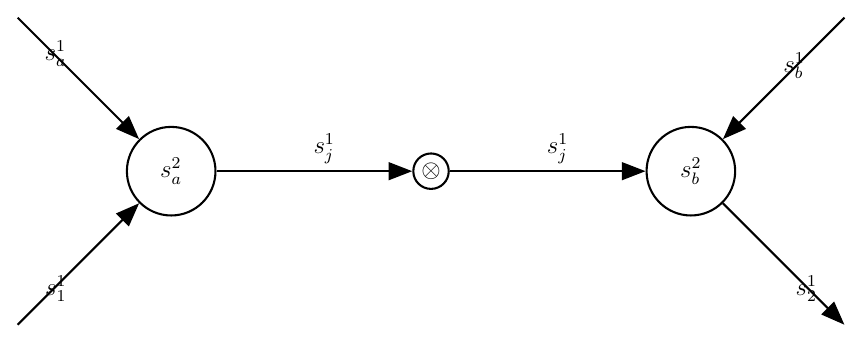}
\par\end{centering}
\caption{One example of a measurement event network, which is the product of two measurement events involving a single particle of spin $j$.}\label{fig: primitive event network Sketch}
\end{figure}

We are now prepared to introduce a convenient graphical representation of the new base-2 picture we have developed thus far. In Figure \ref{fig: all of alices combinations Sketch}, all possible directed graphs associated with Alice's measurement event are depicted. The four possibilities each correspond to one of the allowed positions of the base-2 and base-4 sequences within their respective products. The graph of interest for the physical system being modeled here is in the top left corner. That is, we are interested in a situation in which Alice's reference sequence ($s^1_a$) interacts with a particle ($s^1_1$) producing a new particle ($s^1_j$), which then moves to the right towards Bob's reference sequence ($s^1_b$). As depicted in Figure \ref{fig: primitive event network Sketch}, an interaction then occurs between Bob's reference sequence ($s^1_b$) and the incoming particle ($s^1_j$) producing a new particle ($s^1_2$).

Importantly, all of the allowed directed graphs associated with measurement events must have two edges pointing towards the vertex. As alluded to above, we interpret directed edges as particles and vertices as events. For a directed graph to qualify as a measurement event, we must have two incoming edges (particles), one of which must always be associated with a reference sequence. Of course, this is a logical requirement of measurement. Simply put, only one edge can be directed away from the vertex, and it must never be the one associated with the reference sequence. In a more general treatment of events, which would include events not involving a reference sequence, other graphs may be permitted.

We are now in a position to appreciate the full power of this new conceptual picture. The relative rotation between Alice's and Bob's magnetic field gradients, or the change of measurement basis in the canonical vernacular, can be modeled by the number of differences between their reference sequences. That is, $\theta_{ab}$ is proportional to the Hamming distance between Alice's and Bob's base-2 reference sequences. Thus, the act of rotating field gradients is associated with the following operation:

\begin{equation}
s^1_a \oplus s^1_\theta = s^1_b \qquad s^1_\theta \in S^1(n)  
\label{eq: reference sequence rotation}
\end{equation}

Following from the definition of $\theta_{ab}$ given in equation (\ref{eq: theta}), the relative rotation between Alice's and Bob's reference sequence can be defined in terms of base-2 symbol counts as follows:

\begin{equation}
    \tan\left(\frac{\theta_{ab}}{2}\right)\equiv\frac{\tilde{1}_{\theta}}{\tilde{0}_{\theta}}\label{eq: theta base-2}
\end{equation}

\begin{figure}
\begin{centering}
\includegraphics[scale=.7]{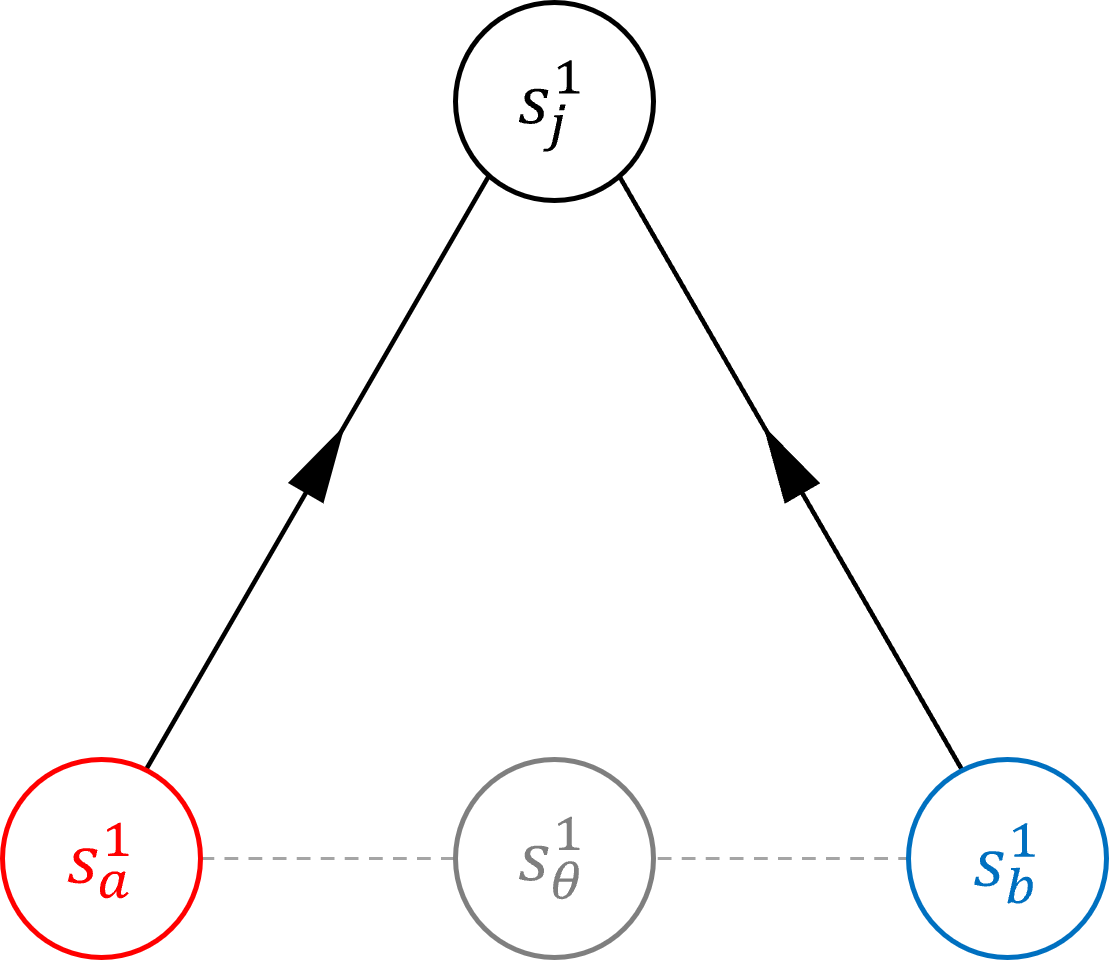}
\par\end{centering}
\caption{Alice and Bob each interact with a common particle, modeled by a base-2 sequence with $2j$ $1$'s. The relative rotation between Alice's and Bob's magnetic field gradients is modeled by the Hamming distance between their reference sequences. As this structure is fully specified by three unique base-2 sequences ($s^1_a\otimes s^1_j \otimes s^1_b$), the full system can be described by a single base-8 sequence.}\label{fig: base-2 model sketch}
\end{figure}

A diagram of the conceptual picture implied by equations (\ref{eq: alices measurement event operation}-\ref{eq: theta base-2}) is provided in Figure \ref{fig: base-2 model sketch}. This picture consists of a network of three unique base-2 sequences ($s^1_a$, $s^1_b$, $s^1_j$), or a single base-8 sequence ($s^1_a\otimes s^1_j \otimes s^1_b$). The magnetic field gradients in Alice's and Bob's detector are each modeled by a base-2 reference sequence with the subscript $a$ or $b$. The relative rotation between these magnetic field gradients is then modeled by the number of $1$'s that appear in the base-2 sequence carrying the subscript $\theta$. This rotation is depicted in equation (\ref{eq: reference sequence rotation}). Likewise, the particle is modeled by a base-2 sequence carrying the subscript $j$, where $2j$ is the number of $1$'s that appear in this sequence. Finally, the interactions between this particle and the magnetic field gradients are modeled by the measurement event operations depicted in equations (\ref{eq: alices measurement event operation}) and (\ref{eq: bobs measurement event operation}). Thus, Alice and Bob each interact with a common particle ontic state ($s^1_j$), where the relative rotation between their magnetic field gradients is modeled by the Hamming distance between their own ontic states ($s^1_a$, $s^1_b$). As will be illustrated at the end of this section, the non-determinism within the proposed model can be attributed solely to a lack of knowledge about the ordering of the base-2 symbols in $s^1_\theta$.

Perhaps the most striking feature of the proposed model is the relational nature of the quantum numbers $m_{a}$ and $m_{b}$, which are properties of the vertices in Figure \ref{fig: primitive event network Sketch}. Unlike $j$, which we may associate with a particle traveling between Alice's and Bob's detectors, the quantum numbers $m_{a}$ and $m_{b}$ are exclusive properties of measurement events. That is, one must know the ontic state of both the particle ($s^1_j$) and the magnetic field gradient ($s^1_a$ or $s^1_b$) to determine $m_{a}$ or $m_{b}$. We can illustrate this important point by holding a base-2 sequence used to model a spin $\frac{1}{2}$ particle fixed ($s^1_j$), while considering all possible reference sequences ($s^1_a$) for $n=2$:

\begin{equation}
\scalebox{0.8}{$\begin{aligned}
  \left(\begin{array}{c}
0\\
0\\
\end{array}\right)_{a}\oplus\left(\begin{array}{c}
0\\
1\\
\end{array}\right)_{j}=\left(\begin{array}{c}
0\\
1\\
\end{array}\right)_{1}
\Rightarrow \left(\begin{array}{c}
A\\
D
\end{array}\right)_{a}
\end{aligned}$}
\label{eq: fix j illustration 1}
\end{equation}

\begin{equation}
\scalebox{0.8}{$\begin{aligned}
  \left(\begin{array}{c}
1\\
0\\
\end{array}\right)_{a'}\oplus\left(\begin{array}{c}
0\\
1\\
\end{array}\right)_{j}=\left(\begin{array}{c}
1\\
1\\
\end{array}\right)_{1}
\Rightarrow \left(\begin{array}{c}
B\\
D
\end{array}\right)_{a'}
\end{aligned}$}
\label{eq: fix j illustration 2}
\end{equation}

\begin{equation}
\scalebox{0.8}{$\begin{aligned}
  \left(\begin{array}{c}
0\\
1\\
\end{array}\right)_{a''}\oplus\left(\begin{array}{c}
0\\
1\\
\end{array}\right)_{j}=\left(\begin{array}{c}
0\\
0\\
\end{array}\right)_{1}
\Rightarrow \left(\begin{array}{c}
A\\
C
\end{array}\right)_{a''}
\end{aligned}$}
\label{eq: fix j illustration 3}
\end{equation}

\begin{equation}
\scalebox{0.8}{$\begin{aligned}
  \left(\begin{array}{c}
1\\
1\\
\end{array}\right)_{a'''}\oplus\left(\begin{array}{c}
0\\
1\\
\end{array}\right)_{j}=\left(\begin{array}{c}
1\\
0\\
\end{array}\right)_{1}
\Rightarrow \left(\begin{array}{c}
B\\
C
\end{array}\right)_{a'''}
\end{aligned}$}
\label{eq: fix j illustration 4}
\end{equation}

By equation (\ref{eq: base-4 quantum numbers}), the value of $m$ associated with the measurement events depicted in equations (\ref{eq: fix j illustration 1}-\ref{eq: fix j illustration 4}) are $-\frac{1}{2}$, $-\frac{1}{2}$, $+\frac{1}{2}$, and $+\frac{1}{2}$, respectively. As claimed, one must have knowledge of the ontic states of both the particle and the magnetic field gradient to determine the quantum number $m$. Thus, $m$ is a relational quantity and an exclusive property of measurement events. Though a full proof is beyond the scope of this paper, it is this feature that allows the model presented here to remain in good standing with Bell's theorem and the Kochen-Specker theorem. This feature also highlights the clarity with which the proposed model accounts for the distribution of information within a quantum system. \textit{Importantly, the information necessary to predict the outcome of a given experiment only becomes localized at the measurement event}. Again, rigorous proofs of these statements are beyond the scope of the work presented here, but the basic elements of these proofs have been established.  

While these are indeed important features for a model of quantum systems, we still have not confirmed explicitly  that our provisional definition of $\theta_{ab}$ yields the correct prediction for the $n=2$ case. Specifically, we are interested in the case of $\theta_{ab}=\frac{\pi}{2}$, which should yield an equal probability of observing $m_{b}=+\frac{1}{2}$ and $m_{b}=-\frac{1}{2}$ at Bob's detector, regardless of what happens at Alice's detector. We begin by choosing an ontic state for Alice's measurement event:

\begin{equation}
    \scalebox{0.7}{$\begin{aligned}
  \left(\begin{array}{c}
0\\
1\\
\end{array}\right)_{a}\otimes\left(\begin{array}{c}
1\\
1\\
\end{array}\right)_{1}=
\left(\begin{array}{c}
D\\
B
\end{array}\right)_{a}
\end{aligned}$}
    \label{eq: Alice's measurement event example}
\end{equation}

We can then determine the ontic state of the spin $\frac{1}{2}$ particle with the following operation:

\begin{equation}
    \scalebox{0.7}{$\begin{aligned}
  \left(\begin{array}{c}
0\\
1\\
\end{array}\right)_{a}\oplus\left(\begin{array}{c}
1\\
1\\
\end{array}\right)_{1}=
\left(\begin{array}{c}
1\\
0
\end{array}\right)_{j}
\end{aligned}$}
    \label{eq: generating the particle ontic state}
\end{equation}

Applying both possible $\theta_{ab}=\frac{\pi}{2}$ rotation maps to Alice's reference sequence yields the following possibilities for Bob's reference sequence: 

\begin{equation}
\scalebox{0.7}{$\begin{aligned}
  \left(\begin{array}{c}
0\\
1\\
\end{array}\right)_{a}\oplus\left(\begin{array}{c}
1\\
0\\
\end{array}\right)_{\theta}=
\left(\begin{array}{c}
1\\
1
\end{array}\right)_{b},
\quad
  \left(\begin{array}{c}
0\\
1\\
\end{array}\right)_{a}\oplus\left(\begin{array}{c}
0\\
1\\
\end{array}\right)_{\theta}=
\left(\begin{array}{c}
0\\
0
\end{array}\right)_{b}
\end{aligned}$}
\label{eq: pi/2 rotation example 1}
\end{equation}

We are left with two possible interactions at Bob's detector, one for each reference sequence:

\begin{equation}
\scalebox{0.7}{$\begin{aligned}
  \left(\begin{array}{c}
1\\
1\\
\end{array}\right)_{b}\oplus\left(\begin{array}{c}
1\\
0\\
\end{array}\right)_{j}=
\left(\begin{array}{c}
0\\
1
\end{array}\right)_{2},
\quad
  \left(\begin{array}{c}
0\\
0\\
\end{array}\right)_{b}\oplus\left(\begin{array}{c}
1\\
0\\
\end{array}\right)_{j}=
\left(\begin{array}{c}
1\\
0
\end{array}\right)_{2}
\end{aligned}$}
\label{eq: generating Bobs measurement event ontic state}
\end{equation}

Finally, we can write down the two possible ontic states associated with Bob's measurement event:

\begin{equation}
\scalebox{0.7}{$\begin{aligned}
  \left(\begin{array}{c}
1\\
1\\
\end{array}\right)_{b}\otimes\left(\begin{array}{c}
0\\
1\\
\end{array}\right)_{2}=
\left(\begin{array}{c}
C\\
B
\end{array}\right)_{b},
\quad
  \left(\begin{array}{c}
0\\
0\\
\end{array}\right)_{b}\otimes\left(\begin{array}{c}
1\\
0\\
\end{array}\right)_{2}=
\left(\begin{array}{c}
D\\
A
\end{array}\right)_{b}
\end{aligned}$}
\label{eq: listing Bobs measurement event ontic state}
\end{equation}

The set of ontic states constructed in equations (\ref{eq: Alice's measurement event example}-\ref{eq: listing Bobs measurement event ontic state}) are the elements of Alice's statistical ensemble, as discussed in section \ref{sec:Changing-variables}. We can write this set explicitly like so:

\begin{equation}
\scalebox{0.8}{$\begin{aligned}
\left(\begin{array}{c}
D\\
B
\end{array}\right)_{a}\otimes\left\{ \left(\begin{array}{c}
C\\
B
\end{array}\right)_{b},\left(\begin{array}{c}
D\\
A
\end{array}\right)_{b}\right\}
\end{aligned}$}
\label{eq: Alice's sample space spin 1/2 example}
\end{equation}

The presence of more than one ontic state in Alice's statistical ensemble can be attributed directly to equation (\ref{eq: pi/2 rotation example 1}), where the two possible $s^1_\theta$ rotation maps were considered. Thus, non-determinism within the proposed model arises by hiding the information stored in the ordering of the base-2 symbols in $s_\theta^1$. Returning to the italicized statement in section (\ref{sec:Changing-variables}): \textit{counts are generally observable, but sequences are not.} Indeed, this statement captures the essence of the procedure outlined in equations (\ref{eq: Alice's measurement event example}-\ref{eq: Alice's sample space spin 1/2 example}). However, the act of holding the base-2 sequences used to construct Alice's measurement event $s^2_{a}$ fixed within the statistical ensemble in equation (\ref{eq: Alice's sample space spin 1/2 example}) appears to contradict this statement. This apparent contradiction can be resolved by noting that the cardinality of Alice's statistical ensemble, which is the physically relevant quantity, is invariant under permutations of Alice's fixed measurement event $s^2_{a}$. In other words, reordering the symbols in $s^2_{a}$ will have no impact on the calculated probabilities. Of course, this will not generally be the case when varying counts. 

Following the Born rule, we must also construct Bob's statistical ensemble to calculate probabilities. To do this, we simply repeat the prescription used to construct Alice's statistical ensemble. That is, we choose a particular ontic state for Bob's measurement event and apply all possible rotation maps that generate Alice's event. The difference is that we know the value of $m_{a}$, but not $m_{b}$. This means that we will actually need to repeat this process twice, once for each possible value of $m_{b}$. One possible example of Bob's statistical ensemble is given here:

\begin{equation}
\scalebox{0.8}{$\begin{aligned}
\left\{ 
\left(\begin{array}{c}
B\\
D
\end{array}\right)_{a}\otimes \left(\begin{array}{c}
B\\
C
\end{array}\right)_{b},\left(\begin{array}{c}
B\\
D
\end{array}\right)_{a}\otimes\left(\begin{array}{c}
A\\
D
\end{array}\right)_{b}\right\}
\end{aligned}$} 
\label{eq: Bob's sample space spin 1/2 example}
\end{equation}

Note that the base-4 sequence used to model Alice's measurement event within her statistical ensemble does not appear in Bob's. Rather, it is a permutation of her measurement event that appears. The point being made here is that the Born rule does not require that Alice's and Bob's statistical ensembles contain common elements. The only quantity of interest when calculating probabilities are cardinalities, or the number of ontic states associated with each unique set of quantum numbers.  

We are finally ready to check if the proposed definition of $\theta_{ab}$ produces the correct prediction for the case of $\theta_{ab}=\frac{\pi}{2}$. The number of times ontic states associated with $m_{b}=+\frac{1}{2}$ occur in Alice's and Bob's statistical ensembles is $1$. Following the Born rule, we count the ways to combine these elements ($1\times1$). The contribution from $m_{b}=-\frac{1}{2}$ is identical ($1\times1$). The final step is to write probabilities in the form of relative frequencies, where the denominator is the sum of these contributions:

\begin{equation}{\begin{aligned}
    P(+\frac{1}{2})=\frac{1\times1}{1\times1+1\times1}=\frac{1}{2}\\
    P(-\frac{1}{2})=\frac{1\times1}{1\times1+1\times1}=\frac{1}{2}\\
    \end{aligned}} 
    \label{eq: first spin 1/2 probability}
\end{equation}

Thus, we have recovered the fact that Bob will observe $+\frac{1}{2}$ and $-\frac{1}{2}$ with equal probability at his detector. Importantly, the non-deterministic nature of this result arises from not specifying a particular ontic state for the rotation map ($s^1_\theta$) relating Alice's and Bob's reference sequences. As previously discussed, non-determinism arises by hiding information stored in the ordering of the symbols comprising sequences. It is also important to note that a specific value of $l_{a}$ was chosen for Alice's measurement event in equation (\ref{eq: Alice's measurement event example}). As mentioned earlier in this section, we will typically sum over all possible values of $l_{a}$ and $l_{b}$ when calculating probabilities for the modeled system. Though, in this case, both possible values of $l_{a}$ lead to the same probability distribution for $m_{b}$. 

Before advancing to the next phase of development, we must reiterate the provisional nature of the conceptual picture presented in this section. The associations made between base-2 sequences and particles, along with the diagrams provided in Figures \ref{fig: all of alices combinations Sketch} and \ref{fig: primitive event network Sketch}, are intended to make these abstract mathematical elements more concrete. By no means should this be viewed as the definitive physical interpretation. Indeed, the correct interpretation of this formalism and model remains an open question.

\section{Spin 1/2  $\quad \theta \in \mathbb{Q}\cap [0,\pi]$ \label{sec: Spin 1/2 arbitrary n}}

The next logical step is to consider arbitrary values of $\theta_{ab}$, which in turn requires $n$ to become arbitrarily large. As we saw in the $n=2$ example, it is essential that we be able to count the number of ontic states associated with a given set of quantum numbers. For the $n=2$ case, this can be done manually due to the relatively small number of ontic states. Unfortunately, as $n$ increases, this approach quickly becomes infeasible. To overcome this, we must borrow some technology from combinatorics. This will enable us to efficiently count the number of sequences associated with each unique set of quantum numbers. However, to make use of this technology, we need to take another step in our formalism. 

At the moment, we can use the quantum numbers $n$, $j$, $m$, and $l$ to count the number of ontic states associated with a given measurement event. We just need a map that converts these quantum numbers into the counts $\tilde{A}$, $\tilde{B}$, $\tilde{C}$, and $\tilde{D}$, which is a straight forward task in linear algebra (Table \ref{tab: base-4}). These counts can then be used in the following combinatorial expression to determine the number of sequences associated with the given quantum numbers, where $n=\tilde{A} + \tilde{B} + \tilde{C} + \tilde{D}$:

\begin{equation}
    \frac{n!}{\tilde{A}!\tilde{B}!\tilde{C}!\tilde{D}!}
    \label{eq: base-4 combinatorial expression}
\end{equation}

Provided we know $n$ and $\theta_{ab}$, we can also use this approach to count the number of possible maps. The only difference in that case is that the counts $\tilde{C}$ and $\tilde{D}$ are always $0$. Thus, given particular choices for the quantum numbers $n$, $j$, $m_{a}$, $l_{a}$, $m_{b}$, $l_{b}$, and $\theta_{ab}$, we can determine the cardinality of the ontic state spaces associated with Alice's and Bob's measurement events, as well as the total number of maps. Unfortunately, this approach to counting does not preserve the full product of Alice's and Bob's measurement events, as depicted in equation (\ref{eq: set of all base-16 sequences}). That is, we will not be able to enforce the full context of the experiment. To do this, we must learn to count base-16 sequences, which will require the introduction of base-16 counts.

As mentioned in section \ref{sec: Spin 1/2 n=2}, we may treat ordered pairs of base-4 sequences as base-16 sequences comprised of the following symbols:

\begin{equation}
\begin{aligned}
    &\left\{AA, AB, AC, AD, BA, BB, BC, BD,\right. \\ 
    &\left. CA, CB, CC, CD, DA, DB, DC, DD
    \right\}
\end{aligned}
    \label{eq: base-16 alphabet}
\end{equation}

For each symbol in this base-16 alphabet, the base-4 symbol on the left is associated with Alice's event and the base-4 symbol on the right is associated with Bob's. Because we are only interested in maps containing the symbols $A$ and $B$, we can actually ignore eight of the symbols introduced in equation (\ref{eq: base-16 alphabet}). This restriction can also be understood as a consequence of Figure \ref{fig: base-2 model sketch}, where the ontic states of the system of interest involve a network of three unique base-2 sequences, rather than four. After accounting for this restriction, we are left with the following set of eight symbols:

\begin{equation}
    \left\{AA, AB, BA, BB, CC, CD, DC, DD
    \right\}
    \label{eq: base-8 alphabet}
\end{equation}

We can now express the base-4 counts associated with Alice's and Bob's events in terms of base-16 counts like so:

\begin{equation}
\tilde{C}_{a}=\widetilde{CC}+\widetilde{CD},\qquad\tilde{D}_{a}=\widetilde{DC}+\widetilde{DD}
\end{equation}
\begin{equation}
\tilde{C}_{b}=\widetilde{CC}+\widetilde{DC},\qquad\tilde{D}_{b}=\widetilde{CD}+\widetilde{DD} 
\end{equation}
\begin{equation}
\tilde{A}_{a}=\widetilde{AA}+\widetilde{AB},\qquad\tilde{B}_{a}=\widetilde{BA}+\widetilde{BB}
\end{equation}
\begin{equation}
\tilde{A}_{b}=\widetilde{AA}+\widetilde{BA},\qquad\tilde{B}_{b}=\widetilde{AB}+\widetilde{BB}
\end{equation}

With these identities, the base-4 quantum numbers introduced thus far become:

\begin{equation}
n=\widetilde{AA}+\widetilde{AB}+\widetilde{BA}+\widetilde{BB}+\widetilde{CC}+\widetilde{CD}+\widetilde{DC}+\widetilde{DD}
\end{equation}

\begin{equation}
j=\frac{1}{2}(\widetilde{CC}+\widetilde{CD}+\widetilde{DC}+\widetilde{DD})
\end{equation}

\begin{equation}
\begin{aligned}
    &m_{a}=\frac{1}{2}(\widetilde{CC}+\widetilde{CD}-\widetilde{DC}-\widetilde{DD}) \\ &m_{b}=\frac{1}{2}(\widetilde{CC}+\widetilde{DC}-\widetilde{CD}-\widetilde{DD})
\end{aligned}
\end{equation}

\begin{equation}
\begin{aligned}
    &l_{a}=\frac{1}{2}(\widetilde{AA}+\widetilde{AB}-\widetilde{BA}-\widetilde{BB}) \\ &l_{b}=\frac{1}{2}(\widetilde{AA}+\widetilde{BA}-\widetilde{AB}-\widetilde{BB})
\end{aligned}
\end{equation}

\begin{equation}
\tan\left(\frac{\theta_{ab}}{2}\right)=\frac{\widetilde{AB}+\widetilde{BA}+\widetilde{CD}+\widetilde{DC}}{\widetilde{AA}+\widetilde{BB}+\widetilde{CC}+\widetilde{DD}}
\end{equation}

So far, we have expressed seven quantum numbers in terms of base-16 counts, each having been previously introduced. However, to count the base-16 sequences of interest in this model, we will need to define an eighth quantum number. This new degree of freedom, which is closely related to the summing parameter q in equation (\ref{eq: Wigner}), will not be important for the spin $\frac{1}{2}$ case. In section \ref{sec: Spin 1 arbitrary n}, however, it will be shown that this degree of freedom is central to the ``quantum" nature of the proposed model. This owes to the fact that, no matter which definition we choose, this final quantum number is an exclusive property of the full ontic state of the experiment. That is, it cannot be associated with Alice's or Bob's measurement events, nor the map which relates the two. There are actually several equivalent ways to define this base-8 quantum number, but the following is perhaps the most instructive:

\begin{equation}
\nu^0_{a,b}\equiv\frac{1}{4}(\widetilde{CD}+\widetilde{DC}+\widetilde{AA}+\widetilde{BB})\label{eq: mu}
\end{equation}

The notation $\nu^0_{a,b}$ has been selected to convey two important points. First, the subscript indicates that this quantity is an exclusive property of the full system being modeled, as previously noted. Second, the superscript indicates that this quantity is just one of a larger set of similar quantities, which become relevant for more complicated systems than the one being modeled here. For example, when treating systems involving networks of four unique base-2 sequences, or all base-16 symbols, there will be six of these special quantum numbers. As these quantities are responsible for driving interference, it is natural to interpret them as path degrees of freedom for the system being modeled. For this reason, we refer to them as path quantum numbers herein.

With all non-zero base-16 quantum numbers defined, we are now able to reintroduce the complete contextual sets $\varepsilon_a$ and $\varepsilon_b$ originally introduced in section \ref{sec: Formalism}:

 \begin{equation}
    \varepsilon_a \equiv \left\{s^4_{a,b}|s^2_{a},n,j,m_{a},m_{b},l_{a},l_{b},\nu^0_{a,b} \right\}
    \label{eq: Formal Alice elementary state space}
\end{equation}

\begin{equation}
    \varepsilon_b \equiv \left\{s^4_{a,b}|s^2_{b},n,j,m_{a},m_{b},l_{a},l_{b},\nu^0_{a,b} \right\}
    \label{eq: Formal Bob elementary state space}
\end{equation}

These contextual sets will be used to construct Alice's and Bob's full statistical ensembles. Before we take this step, however, we must derive the combinatorial expressions necessary to calculate their cardinalities. 
Given a complete set of eight quantum numbers, along with a map which converts these quantum numbers into base-16 counts (Table \ref{tab: base-8}), we can count the base-16 sequences of interest in this model using an analog of equation (\ref{eq: base-4 combinatorial expression}): 

\begin{equation}
    \frac{n!}{\widetilde{AA}!\widetilde{AB}!\widetilde{BA}!\widetilde{BB}!\widetilde{CC}!\widetilde{CD}!\widetilde{DC}!\widetilde{DD}!}
    \label{eq: base-8 combinatorial expression}
\end{equation}

To account for Alice's and Bob's fixed measurement event ontic state in $\varepsilon_a$ and $\varepsilon_b$, we must divide equation (\ref{eq: base-8 combinatorial expression}) by equation (\ref{eq: base-4 combinatorial expression}). This yields the following combinitorial expressions for the cardinality of $\varepsilon_a$ ($|\varepsilon_a|$) and $\varepsilon_b$ ($|\varepsilon_b|$):

\begin{equation}
\begin{aligned}
    |\varepsilon_a(n,j,m_{a},m_{b},l_{a},l_{b},\theta_{ab},\nu^0_{a,b})|=& \\
    \frac{\tilde{A}_{a}!\tilde{B}_{a}!\tilde{C}_{a}!\tilde{D}_{a}!}{\widetilde{AA}!\widetilde{AB}!\widetilde{BA}!\widetilde{BB}!\widetilde{CC}!\widetilde{CD}!\widetilde{DC}!\widetilde{DD}!}&
\end{aligned}
    \label{eq: varepsilon^a cardinality}
\end{equation}

\begin{equation}
    \begin{aligned}
    |\varepsilon_b(n,j,m_{a},m_{b},l_{a},l_{b},\theta_{ab},\nu^0_{a,b})|=& \\
    \frac{\tilde{A}_{b}!\tilde{B}_{b}!\tilde{C}_{b}!\tilde{D}_{b}!}{\widetilde{AA}!\widetilde{AB}!\widetilde{BA}!\widetilde{BB}!\widetilde{CC}!\widetilde{CD}!\widetilde{DC}!\widetilde{DD}!}&
    \end{aligned}
    \label{eq: varepsilon^b cardinality}
\end{equation} 

Before taking the next step in calculating probabilities, we point out the similarity between equations (\ref{eq: alice base-4 counts new}-\ref{eq: bob base-4 count new}) and (\ref{eq: varepsilon^a cardinality}-\ref{eq: varepsilon^b cardinality}). Clearly, we have recovered the combinatorial terms associated with the symbols $C$ and $D$, but with the addition of new terms involving the symbols $A$ and $B$. These terms, along with the normalization scheme employed at the end of section \ref{sec: Spin 1/2 n=2}, play an analogous role to the sine and cosine terms in equation (\ref{eq: Wigner}).

We are nearly ready to construct Alice's and Bob's full statistical ensembles and calculate probabilities for arbitrary rotations. To aid in this final stage, it will be helpful to formally categorize all eight quantum numbers as either random variables ($R$), conditioning variables ($U$), nuisance variables ($W$), or path variables ($\Lambda$). In this case, there is only one random variable, which is $(m_{b})=R$. The conditioning variables, which are the fixed variables in the physical system of interest, are $(n,j,m_{a},\theta_{ab})=U$. The nuisance variables will generally be $(l_{a},l_{b})=W$, but treating $l_{a}$ as a conditioning variable has an interesting effect on the resulting probability distribution (Figure \ref{fig: spin 1/2 plot}). Finally, the only path variable will be ($\nu^0_{a,b})=\Lambda$, which will actually be a constant for the spin $\frac{1}{2}$ case.

We are now ready to construct Alice's and Bob's full statistical ensembles for the system of interest. Though, it will be instructive to break up this construction into two steps. In the first step, we form a union of Alice's and Bob's contextual sets ($\varepsilon_a$ and $\varepsilon_b$) over the path quantum number $\nu^0_{a,b}$. We call these intermediate ensembles ``local" ensembles. Of course, these local ensembles are rather trivial for the spin $\frac{1}{2}$ case, where only one value of $\nu^0_{a,b}$ is possible for each choice of the other seven quantum numbers. For higher order spin systems, however, these intermediate ensembles will become very important. Specifically, they will play a central role in the phenomenon of interference. These local ensembles are defined as follows, where $Q=R\cup U\cup W \cup \Lambda$ is the set of all possible combinations of quantum numbers and we use the N-tuples $r \in R$, $u \in U$, $w \in W$, and $\lambda \in \Lambda$ as shorthand for the eight quantum numbers:

\begin{equation}
    \qquad L_a(r,u,w) \equiv \bigcup_{\lambda\in Q(r,u,w)}\varepsilon_a(r,u,w,\lambda)
    \label{eq: Alice's local ensemble}
\end{equation}

\begin{equation}
    L_b(r,u,w) \equiv \bigcup_{\lambda\in Q(r,u,w)}\varepsilon_b(r,u,w,\lambda)
    \label{eq: Bob's local ensemble}
\end{equation}

The set $Q$ is critical within this model. Not only in the construction of Alice's and Bob's statistical ensembles, but also in the calculation of probabilities. Note that it can be found by solving the system of equations given in Table \ref{tab: base-8}. Though this set grows in size quickly, it will always be finite for finite $n$. An alternative approach to defining these ensembles, which does not explicitly involve the set $Q$, is as follows:

\begin{equation}
L_a(r,u,w) \equiv \left\{s^4_{a,b}|s^2_{a},r,u,w \right\}
\label{eq: alt Alice's local ensemble}
\end{equation}

\begin{equation}
L_b(r,u,w) \equiv \left\{s^4_{a,b}|s^2_{b},r,u,w \right\}
\label{eq: alt Bob's local ensemble}
\end{equation}

By construction, the path quantum numbers $\lambda$ are symmetries of $L_a$ and $L_b$. In the following section, we will discuss the symmetry operations that relate the ontic states within $L_a$ and $L_b$ in some detail. Importantly, these symmetry operations will always lead to integer variations in the path quantum numbers. Given some initial choice of $\lambda$, which we denote as $\lambda_0$, the total variation in $\lambda$ due to an arbitrary sequence of symmetry operations is as follows:

\begin{equation}
    \Delta \lambda(\lambda,\lambda_0) \equiv  \sum_{i}\lambda^i - \lambda^i_0
\end{equation}

When calculating probabilities, the ontic states associated with even values of $\Delta \lambda$ will interfere destructively with ontic states associated with odd values of $\Delta \lambda$. The relevant quantities are the following, where we again use the $r$, $u$, $w$, and $\lambda$ shorthand and require that $\lambda_0$ be the same for both expressions:

\begin{equation}
\begin{aligned}[t]
&\Upsilon_a(r,u,w)\equiv \sum_{\lambda \in Q(r,u,w)}(-1)^{\Delta \lambda(\lambda,\lambda_0)}|\varepsilon_a(r,u,w,\lambda)|\!\!\!\!\!\!\!\!
\end{aligned}
\label{eq:  upsilon a}
\end{equation}

\begin{equation}
\begin{aligned}[t]
&\Upsilon_b(r,u,w)\equiv \sum_{\lambda\in Q(r,u,w)}(-1)^{\Delta \lambda(\lambda,\lambda_0)}|\varepsilon_{b}(r,u,w,\lambda)|\!\!\!\!\!\!\!
\end{aligned}
\label{eq:  upsilon b}
\end{equation}

We are now ready to construct Alice's and Bob's full statistical ensembles for the system of interest:

\begin{equation}
    E_a(u) \equiv \bigcup_{(r,w)\in Q(u)} L_a(r,u,w)
    \label{eq: Alice's statistical ensemble}
\end{equation}

\begin{equation}
    E_b(u) \equiv \bigcup_{(r,w)\in Q(u)}L_b(r,u,w)
    \label{eq: Bob's statistical ensemble}
\end{equation}

Finally, we can use the expressions defined in equations (\ref{eq:  upsilon a}) and (\ref{eq:  upsilon b}) to build our general expression for probabilities. Given $u$, the probability of observing a particular value of $r$ is the product of equation (\ref{eq: upsilon a}) and (\ref{eq:  upsilon b}) summed over $w$ and normalized by the contributions from all possible values of $r$:

\begin{equation}
    P(r|u)=
    \frac{\sum_{w\in Q(r,u)}\Upsilon_a(r,u,w) \Upsilon_b(r,u,w)}{\sum_{(r,w) \in Q(u)}\Upsilon_a(r,u,w) \Upsilon_b(r,u,w)}
\end{equation}

This expression, which is identical to the one in equation (\ref{eq: final}) is compared to equation (\ref{eq: Wigner}) in Figure \ref{fig: spin 1/2 plot} for the spin $\frac{1}{2}$ case. Under the calibration (\ref{eq: theta}) the two curves are indistinguishable at the resolution of the plot, with the residual peaking near $\frac{\pi}{4}$ and $\frac{3\pi}{4}$ at approximately $2.5\times10^{-3}$ and vanishing at $\theta_{ab}=0$, $\frac{\pi}{2}$ and $\pi$. The nuisance variable $l_{a}$ is summed over throughout; it is not used as a tuning parameter. The rate at which QM is recovered is shown in Figure \ref{fig: spin 1/2 plot} (bottom), where the maximum of $|\Delta|$ over $\theta_{ab}$ is plotted against $n$ for each of the three systems treated in this paper. In every case it falls as $O(1/n)$, halving with each doubling of $n$, so that QM is recovered exactly in the limit of large $n$. Representative values are $2.50\times10^{-3}$, $1.25\times10^{-3}$ and $6.2\times10^{-4}$ at $n=100$, $200$ and $400$ for $j=\frac{1}{2}$, and $1.02\times10^{-2}$, $5.08\times10^{-3}$ and $2.54\times10^{-3}$ for $j=1$ with $m_{a}=+1$. It is worth recording what the earlier linear assignment $\theta_{ab}=(\tilde{B}_{map}/n)\pi$ gives by comparison: there the same quantity converges instead to a non-zero constant, approaching $0.050$ for $j=\frac{1}{2}$, $0.080$ for $j=1$ with $m_{a}=+1$, and $0.137$ for $j=1$ with $m_{a}=0$. That residual is a genuine disagreement with QM which no amount of continuity removes, and it is what previously had to be absorbed by treating $l_{a}$ as a tuning parameter. With the correct calibration no such device is required.

There may be other ways to refine the model proposed here.
This is made possible by the first principles construction of the
expression in equation (\ref{eq: final}). At each step in the
development of that expression, certain choices were made that could
be modified. For example, we could have chosen a different set of
quantum numbers. Perhaps some alternative to $l_{a}$ and $l_{b}$
would produce a better result. Modifications of this type are of interest for future work. Though, the deviations observed in Figure \ref{fig: spin 1/2 plot} vanish in the limit of large $n$, and at finite $n$ they are not of the type that lead to violations of no-go theorems.

\section{Spin 1  $\quad \theta \in \mathbb{Q}\cap [0,\pi]$ \label{sec: Spin 1 arbitrary n}}

\begin{figure}
\begin{centering}
\includegraphics[scale=.65]{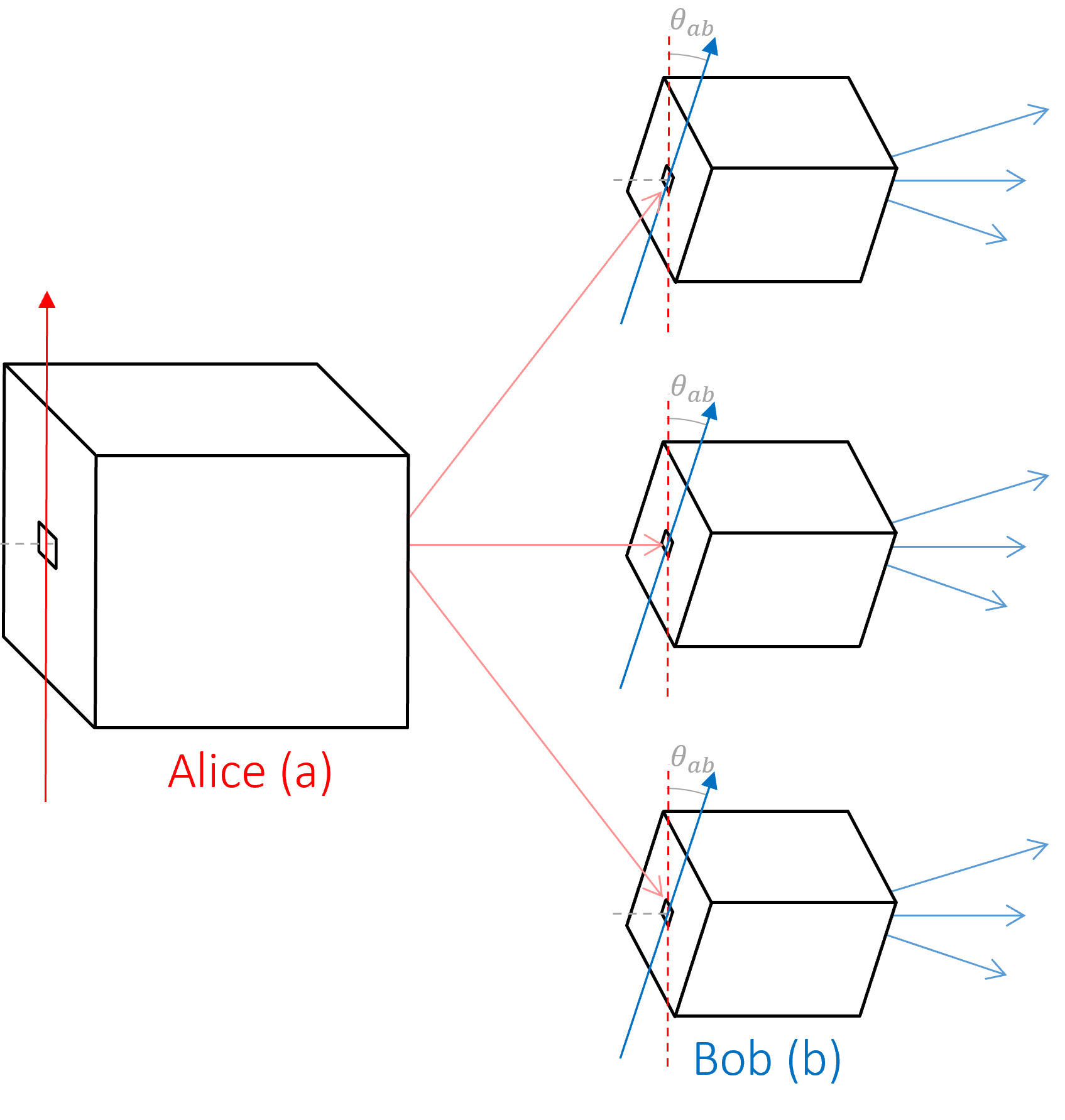}
\par\end{centering}
\caption{A measurement event Alice's detector deflects the $j=1$ particle into one of three paths (red), one for each possible value of the quantum number $m_{a}\in \{+1,0,-1\}$. Depending on which value of $m_{a}$ is of interest, Bob's detector is then placed along one of these paths. Bob then rotates his detector with respect to Alice's by the angle $\theta_{ab}$. Finally, a measurement event at Bob's detector deflects the $j=1$ particle into one of three paths (blue), one for each possible value of the quantum number $m_{b}\in \{+1,0,-1\}$.}\label{fig: sketch 3}
\end{figure}

In this section, we study the phenomenon of interference within the proposed model. As discussed in the previous section, the simplest system for which this becomes relevant is spin $1$, as illustrated in Figure \ref{fig: sketch 3}. A comparison of equations (\ref{eq: final}) and (\ref{eq: Wigner}) for this system is provided in Figure \ref{fig: j=1}, where we have treated $l_{a}$ as a nuisance variable. The case of interest for this section is the solid purple line in the bottom plot, which corresponds to $m_{a}=m_{b}=0$. As $\theta_{ab}$ approaches $\frac{\pi}{2}$, the probability that Alice and Bob both observe $m=0$ vanishes. This is a direct consequence of interference. 

In equations (\ref{eq: alt Alice's local ensemble}) and (\ref{eq: alt Bob's local ensemble}), the local ensembles $L_a$ and $L_b$ were defined in a particularly instructive manner. In $L_a$, each ontic state shares the same base-4 sequence $s^2_{a}$, corresponding to a fixed choice of Alice's measurement event. One possible element of $L_a$ is offered below, where $n=6$, $j=1$, $m_{a}=m_{b}=0$, $l_{a}=0$, $l_{b}=+1$, $\theta_{ab}=\frac{\pi}{2}$, and $\nu^0_{a,b}=\frac{5}{4}$:

\begin{equation}
\scalebox{.8}{$\begin{aligned}
\left(\begin{array}{c}
AA\\
BB\\
DC\\
CD\\
BA\\
AA
\end{array}\right)\end{aligned}$}
\label{eq: nu example}
\end{equation}

As mentioned in the previous section, path quantum numbers are symmetries of the local ensembles $L_a$ and $L_b$. This implies that a symmetry operation exists within $L_a$, for example, that generates variations in $\nu^0_{a,b}$ while leaving Alice's measurement event ($s^2_{a}$) and all remaining quantum numbers ($n,j,m_{a},m_{b},l_{a},l_{b},\theta_{ab}$) invariant. This operation can be expressed in terms of base-4 transposition operators, which exchange the position of any two symbols within a base-4 sequence. The operators are denoted as follows, where the superscript indicates the symbols being transposed and the subscript indicates the base-4 sequence being operated on:

\begin{equation}
    T^{AB}_{a}, T^{AC}_{a}, T^{AD}_{a}, T^{BC}_{a}, T^{BD}_{a}, T^{CD}_{a}
    \label{eq: a1 operators}
\end{equation}

\begin{equation}
    T^{AB}_{b}, T^{AC}_{b}, T^{AD}_{b}, T^{BC}_{b}, T^{BD}_{b}, T^{CD}_{b}
    \label{eq: b2 operators}
\end{equation}

The operators of interest when studying the symmetries of $L_a$ are those in equation (\ref{eq: b2 operators}), which generate variations in Bob's measurement event ontic state. While these transposition operators will always leave the base-4 quantum numbers associated with Alice's and Bob's measurement event invariant ($n,j,m_{a},m_{b},l_{a},l_{b}$), they will not necessarily leave $\theta_{ab}$ invariant. One example of this can be seen here: 

\begin{equation}
\scalebox{0.8}{$\begin{aligned}
\left(\begin{array}{c}
\textbf{AA}\\
\textbf{BB}\\
DC\\
CD\\
BA\\
AA
\end{array}\right)\xrightarrow{T^{AB}_{b}}\left(\begin{array}{c}
\textbf{AB}\\
\textbf{BA}\\
DC\\
CD\\
BA\\
AA
\end{array}\right)
\end{aligned}$}
\label{eq: transpostion 1}
\end{equation}

In this example, the transposition operator caused $\theta_{ab}$ to change from $\frac{\pi}{2}$ to $\frac{5\pi}{6}$. That is, the operation depicted in equation (\ref{eq: transpostion 1}) is not a symmetry of $L_a$. To generate the desired symmetry, a given transposition operator must act on the correct collection of base-16 symbols, or operands. For example, the operands $(AA,AB)$, $(BA,BB)$, $(CA,CB)$, and $(DA,DB)$ will result in a symmetry of $L_a$ when acted upon by the transposition operator $T^{AB}_{b}$. One such example is offered here, where $T^{AB}_{b}$ acts on $(BA,BB)$:

\begin{equation}
\scalebox{0.8}{$\begin{aligned}
\left(\begin{array}{c}
AA\\
\textbf{BB}\\
DC\\
CD\\
\textbf{BA}\\
AA
\end{array}\right)\xrightarrow{T^{AB}_{b}}\left(\begin{array}{c}
AA\\
\textbf{BA}\\
DC\\
CD\\
\textbf{BB}\\
AA
\end{array}\right)
\end{aligned}$}
\label{eq: transpostion 2}
\end{equation}

The operation depicted in equation (\ref{eq: transpostion 2}) actually leaves all eight quantum numbers invariant. This implies that, in addition to being a symmetry of $L_a(r,u,w)$, it is also a symmetry of the contextual set $\varepsilon_a(r,u,w,\lambda)$. In other words, it does not generate a variation in the path quantum number, as desired. To generate these variations, we must use more than one transposition operator. In the case of $\nu^0_{a,b}$, the pair of interest is $(T^{AB}_{b},T^{CD}_{b})$, which may act on the operands $(AB,BA,CC,CD)$ or $(AA,BB,CD,DC)$ to generate the desired symmetry. Examples of each of these operations are given below:   

\begin{equation}
\scalebox{0.8}{$\begin{aligned}
\left(\begin{array}{c}
\textbf{AB}\\
\textbf{BA}\\
\textbf{DD}\\
\textbf{CC}\\
BA\\
AA
\end{array}\right)_{\nu^0_{a,b}=\frac{1}{4}}
\xrightarrow{T^{AB}_{b}T^{CD}_{b}}\left(\begin{array}{c}
\textbf{AA}\\
\textbf{BB}\\
\textbf{DC}\\
\textbf{CD}\\
BA\\
AA
\end{array}\right)_{\nu^0_{a,b}=\frac{5}{4}}
\end{aligned}$}
\label{eq: transpostion 3}
\end{equation}

\begin{equation}
\scalebox{0.8}{$\begin{aligned}
\left(\begin{array}{c}
AA\\
\textbf{BB}\\
\textbf{DC}\\
\textbf{CD}\\
BA\\
\textbf{AA}
\end{array}\right)_{\nu^0_{a,b}=\frac{5}{4}}
\xrightarrow{T^{AB}_{b}T^{CD}_{b}}\left(\begin{array}{c}
AA\\
\textbf{BA}\\
\textbf{DD}\\
\textbf{CC}\\
BA\\
\textbf{AB}
\end{array}\right)_{\nu^0_{a,b}=\frac{1}{4}}
\end{aligned}$}
\label{eq: transpostion 4}
\end{equation}

As indicated by the subscripts, these operations indeed generate integer variations in the path quantum number $\nu^0_{a,b}$. In equation (\ref{eq: transpostion 3}), $(T^{AB}_{b},T^{CD}_{b})$ acted on the operand $(AB,BA,CC,DD)$, generating $\Delta\lambda=+1$. In equation (\ref{eq: transpostion 4}), $(T^{AB}_{b},T^{CD}_{b})$ acted on the operand $(AA,BB,CD,DC)$, generating $\Delta\lambda=-1$. Thus, the transposition operator pair $(T^{AB}_{b},T^{CD}_{b})$ can be thought of as a ladder operator, raising or lowering $\nu^0_{a,b}$ depending on which operand is acted upon. When calculating the terms defined in equations (\ref{eq:  upsilon a}) and (\ref{eq:  upsilon b}), these ontic states will interfere destructively due to the interference term $(-1)^{\Delta\lambda}$. 

A complete discussion of $L_a$ and $L_b$ symmetry operations is beyond the scope of this work. However, it should be understood that the special case considered here is only a small part of a large family of symmetry operations. These additional operations become important when modeling systems involving all base-16 symbols, rather than the 8 required for the system of interest here. Specifically, the operator pair $(T^{AB}_{b},T^{CD}_{b})$ is just one of 27 (3 commutative and 24 non-commutative) that lead to variations in a complete set of 6 path quantum numbers. A more thorough analysis of this rich structure is of considerable interest for future work.

\section{Testing the model \label{sec:Testing-the-model}}

\begin{figure}
  \centering
  \includegraphics[width=0.85\columnwidth]{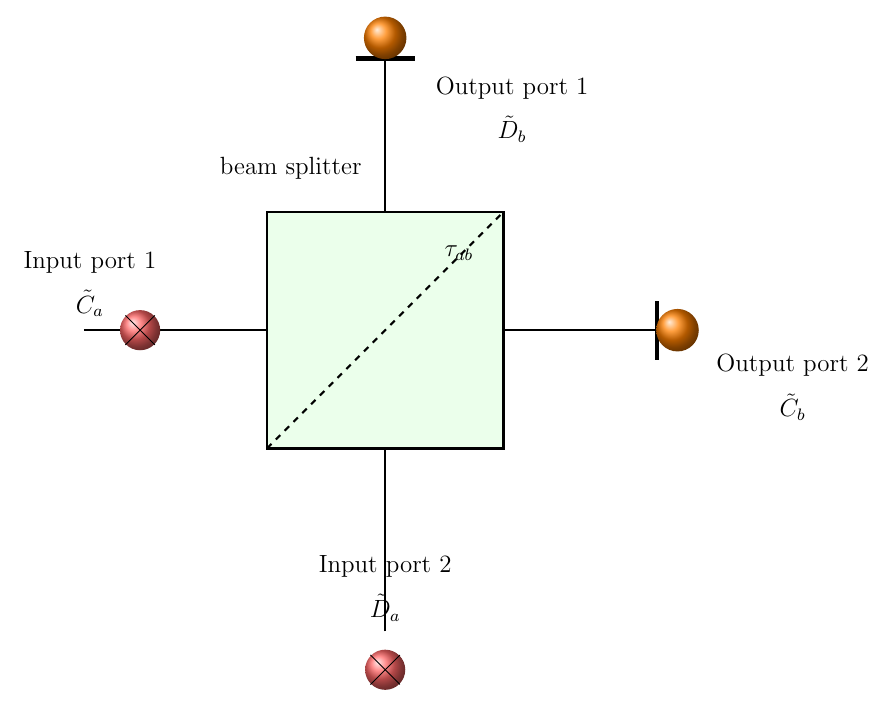}
  \hspace{1cm}
  \includegraphics[scale=1]{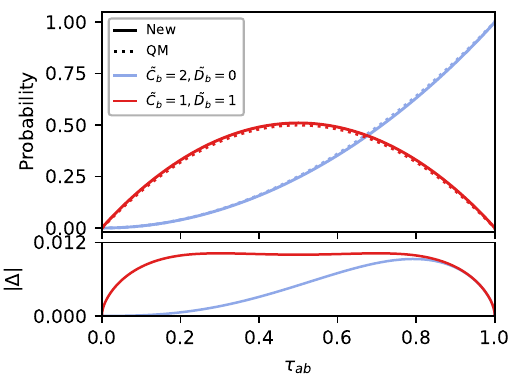}
  \caption{Top: Photon number states modeled by the base-4 counts $\tilde{C}_{a}$ and $\tilde{D}_{a}$ enter the input ports of a beam splitter with transmittance $\tau_{ab}$. The photon number states exiting the output ports of the beam splitter, which are modeled by the base-4 counts $\tilde{C}_{b}$ and $\tilde{D}_{b}$, then interact with a pair of detectors. Bottom: (blue) Probability of $\tilde{C}_{b}=2$, $\tilde{D}_{b}=0$ given
$\tilde{C}_{a}=2$, $\tilde{D}_{a}=0$. (red) Probability of
$\tilde{C}_{b}=1$, $\tilde{D}_{b}=1$ given $\tilde{C}_{a}=2$,
$\tilde{D}_{a}=0$. In all cases, $n=100$ and $|\Delta|$ is the magnitude of the difference
between the new model and QM.}
\label{fig: beam splitter sketch and transmittance}
\end{figure}

With just a few modifications, the model developed for sequences of
SG detectors can also be used to model the behavior of photon number
states passing through a beam splitter \citep{Campos1989}. This is particularly important for recent advances in quantum algorithms for higher order Fock states utilizing variable beam splitters \citep{Sturges_2021}, where our results could lead to significant changes. The photon number state entering the two input ports of a beam splitter are modeled using the base-4 counts $\tilde{C}_{a}$ and $\tilde{D}_{a}$,
while the photon number states exiting the two output ports are modeled using $\tilde{C}_{b}$ and $\tilde{D}_{b}$  (see Figure \ref{fig: beam splitter sketch and transmittance} (top)). 

The effect of the beam splitter is then modeled using a map composed
of $A$ and $B$ symbols only, where the following definition relates
the number of $B$'s within these maps to the transmittance ($\tau_{ab}$)
of the beam splitter:

\begin{equation}
\tau_{ab}\equiv \cos^{2}\left(\frac{\theta_{ab}}{2}\right)=\frac{\tilde{A}_{map}^{2}}{\tilde{A}_{map}^{2}+\tilde{B}_{map}^{2}}\label{eq: transmittance}
\end{equation}

The only remaining modifications necessary are the base-4 quantum numbers
we defined in section \ref{sec: Formalism}. Instead of using $j$,
$m_{a}$, and $m_{b}$, we will use the following, where $\tilde{D}_{b}=\tilde{C}_{a}+\tilde{D}_{a}-\tilde{C}_{b}$:

\begin{equation}
\tilde{C}_{a}=\widetilde{CC}+\widetilde{CD}\label{eq: photon number in 1}
\end{equation}
\begin{equation}
\tilde{D}_{a}=\widetilde{DD}+\widetilde{DC}\label{eq: photon number in 2}
\end{equation}
\begin{equation}
\tilde{C}_{b}=\widetilde{CC}+\widetilde{DC}\label{eq: photon number out 1}
\end{equation}

To calculate probabilities, we simply break the eight quantum numbers into random, conditioning, nuisance, and path variables and use equation (\ref{eq: final}):

\begin{equation}
\begin{aligned}
    (\tilde{C}_{b})=R, \quad (n,\tilde{C}&_{a},\tilde{D}_{a},\tau_{ab})=U,\\ (l_{a},l_{b})=W, &\quad (\nu^0_{a,b})=\Lambda
\end{aligned}
\end{equation}

Through a superficial change of variables, we have applied
the model for sequences of SG detectors to a new physical system.
This particular application is noteworthy because of the experimental
advantages we gain by working with optical systems, rather than spin
systems. This makes high precision tests of the proposed model more
practical and cost effective \citep{Thomay2017}. In particular, the proposed configuration avoids some experimental pitfalls, such as lossy beam splitters or non-ideal detectors \citep{Eisaman2011, Bienfang2023}.

In Figure \ref{fig: beam splitter sketch and transmittance} (bottom),
we offer a comparison of equation (\ref{eq: final})
with QM for two different combinations of $\tilde{C}_{a}$, $\tilde{D}_{a}$,
$\tilde{C}_{b}$, and $\tilde{D}_{b}$ as a function of $\tau_{ab}$,
where $n=100$. To mitigate experimental challenges such as determining the beam splitter splitting ratio $\tau_{ab}$ for a wide variety of values, a comparative measurement on the same beam splitter would be preferable. For example, in Figure \ref{fig: beam splitter sketch and transmittance} (bottom) a specific beam splitter value such as $\tau_{ab}=0.4$ may be used, where at $n=100$ the new model departs from QM by different amounts for the two output states shown. We note that, unlike under the earlier linear assignment, these departures are not fixed features of the model: they fall as $O(1/n)$, the maximum of $|\Delta|$ over $\tau_{ab}$ decreasing from $4.1\times10^{-2}$ at $n=25$ to $1.0\times10^{-2}$ at $n=100$ and $5.1\times10^{-3}$ at $n=200$. With this approach, we can use a specific input configuration ($\widetilde C_{a}=2$,  $\widetilde D_{a}=0$) and output configuration ($\widetilde C_{b}=2$,  $\widetilde D_{b}=0$) as a calibration for this specific beam splitter ratio. For the same input configuration ($\widetilde C_{a}=2$,  $\widetilde D_{a}=0$), but different output configuration ($\widetilde C_{b}=1$,  $\widetilde D_{b}=1$), we would be able to identify any experimental discrepancies for the specific beam splitter ratio. Using this comparative approach practically eliminates any experimentally hard to determine parameters.

As discussed at the end of section \ref{sec: Spin 1/2 arbitrary n}, agreement between the new model and QM improves as $O(1/n)$, so experiments like the one we have suggested provide an opportunity to place a lower bound on $n$: an experiment which resolves probabilities at the level $\delta$ and observes no deviation requires $n\gtrsim O(1/\delta)$. However, small deviations from QM are unavoidable for finite $n$. This arises due to the definition of $\theta_{ab}$ (or $\tau_{ab}$) within the proposed model. In QM, this quantity is assumed to be an element of the real number line $\mathbb{R}$. In the new model, it is an element of the rational number line $\mathbb{Q}$ when $n$ is finite. This granularity implies that for $\theta_{ab}$ ($\tau_{ab}$) sufficiently close to $\pi$ ($0$) or $0$ ($1$), certain combinations of $m_{a}$ and $m_{b}$ ($\tilde{C}_{a}$, $\tilde{D}_{a}$, $\tilde{C}_{b}$, $\tilde{D}_{b}$) will not be possible in the new model, but are possible in QM. The following is a simple example, where $n=6$, $j=1$, $m_{a}=+1$ and $\tilde{B}_{map}=1$, which by equation (\ref{eq: theta}) corresponds to $\theta_{ab}=2\arctan(1/5)\approx0.395$:

\begin{equation}
\scalebox{0.8}{$\begin{aligned}
\left(\begin{array}{c}
A\\
C\\
B\\
C\\
B\\
A
\end{array}\right)_{a}\oplus\left(\begin{array}{c}
A\\
A\\
A\\
B\\
A\\
A
\end{array}\right)_{map}=\left(\begin{array}{c}
A\\
C\\
B\\
D\\
B\\
A
\end{array}\right)_{b}\rightarrow(m_{b}=0)
\end{aligned}$}
\label{eq: discrete theta 1}
\end{equation}

\begin{equation}
\scalebox{0.8}{$\begin{aligned}
\left(\begin{array}{c}
A\\
C\\
B\\
C\\
B\\
A
\end{array}\right)_{a}\oplus\left(\begin{array}{c}
A\\
A\\
A\\
A\\
B\\
A
\end{array}\right)_{map}=\left(\begin{array}{c}
A\\
C\\
B\\
C\\
A\\
A
\end{array}\right)_{b}\rightarrow(m_{b}=+1)
\end{aligned}$}
\label{eq: discrete theta 2}
\end{equation}

The missing spin state in equations (\ref{eq: discrete theta 1})
and (\ref{eq: discrete theta 2}) is $m_{b}=-1$.
Because there is only one $B$ in the map relating Alice's and Bob's
event, there is no way to generate this state. So, the probability
of observing $m_{b}=-1$ is zero according
to the new model, where as it is $\approx0.15\%$ in QM. The magnitude
of this signal will decrease with increasing $n$, however. If we
continue with $\tilde{B}_{map}=1$, but increase $n$ to $60$, the
probability predicted by QM becomes $\approx8\times10^{-6}\%$. Though this
signal becomes increasingly difficult to detect for large $n$, the
fact that the null condition indicates discovery may provide an experimental
advantage.

\section{Discussion \label{sec: Discussion}}

Though many open questions remain, the similarity between the predictions generated by the model presented here and canonical QM brings some credence to the proposed conceptual picture, the central figures of which are measurement event networks, observers, and statistical ensembles. We argue that many important features of quantum theory, such as non-determinism, non-locality, and contextuality, become more clear in this new view. While a detailed treatment of each of these is beyond the scope of this paper, we offer here a short comment on each. 

Non-determinism is an empirical property of nature. Yet, the physical
origins of this property remain unclear. A great deal of debate on
this and related issues have taken place within the context of QM.
In particular, there is the question of whether or not the wavefunction
is itself ontic, or if it results from incomplete information \cite{Leifer2014}.
Because a one-to-one correspondence between the wavefunction in QM
and the statistical ensembles constructed in section \ref{sec: Spin 1/2 arbitrary n} is lacking, framing the new model within the context of this debate is
challenging. With that being said, there is no question that probabilities are epistemic
within this model. That is, they arise due to observers' inability
to resolve certain details about the ontic state of the physical system
under study. For the time being, we remain agnostic with respect to
the physical interpretation of this obscurement, though we do argue
that the underlying ontic state indeed exists. Importantly, the proposed ontology consists of event networks, rather than particles, waves, or fields. 

Of course, a wide variety of epistemic models are possible, but two critical
features distinguish the one presented here. First, the quantum numbers
associated with measurement events are functions of the observer's reference sequence, making them inherently relational \cite{Rovelli1996}. This implies that the information associated with a given measurement event only becomes local at the measurement event itself. This feature is a critical element of models consistent with Bell's theorem and the Kochen-Specker theorem \cite{Bell1964, Clauser1969, Kochen1968}. Furthermore, path quantum numbers become relevant for spin 1 systems and higher, distinguishing them from spin $\frac{1}{2}$ systems in tests of the Kochen-Specker theorem. The second feature of note is the disjointness of quantum states within this model. That is, no single ontic state can ever be associated with more than one unique combination of quantum numbers. This feature distinguishes this model from the well known toy model by Spekkens, for example \cite{Spekkens2007}. It also negates the applicability of the PBR no-go theorem \cite{Pusey2012}. 

Beyond the narrow set of issues we have briefly discussed, there remain
many open questions about the proposed model and the underlying formalism.
These questions can only be addressed through continued model development
and testing. Of particular interest is a model for a Bell test, where preliminary results indicate a clear violation Bell's inequalities \cite{powers2023}. Extending the model developed in this paper to include a third SG detector is also of interest, which would enable us to explore the issues of non-commutativity, Heisenberg's Uncertainty Principle, and the possibility of modeling additional spacetime degrees of freedom. We are hopeful that this fresh perspective on fundamental physical systems will reinvigorate interest in the foundations of quantum theory as a pathway to discovery.

\section*{Acknowledgments}
We would like to thank Djordje Minic, Tatsu Takeuchi, Lauren Hay, and Omar Elsherif for comments on previous versions of this work as well as many helpful discussions. D.S. is partially supported by the US National Science Foundation, under Grant no. PHY-2014021.

\appendix

\section{Tables}

\begin{table}[h]
\begin{centering}
\setlength{\tabcolsep}{12pt}
\begin{tabular}{ll}
$\tilde{A}_{a}=\frac{n}{2}-j+l_{a}$ & $\tilde{A}_{b}=\frac{n}{2}-j+l_{b}$\tabularnewline
$\tilde{B}_{a}=\frac{n}{2}-j-l_{a}$ & $\tilde{B}_{b}=\frac{n}{2}-j-l_{b}$\tabularnewline
$\tilde{C}_{a}=j+m_{a}$ & $\tilde{C}_{b}=j+m_{b}$\tabularnewline
$\tilde{D}_{a}=j-m_{a}$ & $\tilde{D}_{b}=j-m_{b}$\tabularnewline
$\tilde{A}_{map}=\frac{n}{1+\tan(\theta_{ab}/2)}$ & $\tilde{B}_{map}=\frac{n\tan(\theta_{ab}/2)}{1+\tan(\theta_{ab}/2)}$\tabularnewline
\end{tabular}
\par\end{centering}
\caption{Base-4 counts as functions of quantum numbers for the model of interest \label{tab: base-4}}
\end{table}

\begin{table}[h]
\begin{centering}
\begin{tabular}{l}
$\widetilde{AA}=\frac{1}{4}\tilde{A}_{map}-\frac{1}{2}j+\frac{1}{2}l_{a}+\frac{1}{2}l_{b}+\nu^0_{a,b}$\tabularnewline
$\widetilde{AB}=\frac{1}{4}\left(n+\tilde{B}_{map}\right)-\frac{1}{2}j+\frac{1}{2}l_{a}-\frac{1}{2}l_{b}-\nu^0_{a,b}$\tabularnewline
$\widetilde{BA}=\frac{1}{4}\left(n+\tilde{B}_{map}\right)-\frac{1}{2}j-\frac{1}{2}l_{a}+\frac{1}{2}l_{b}-\nu^0_{a,b}$\tabularnewline
$\widetilde{BB}=\frac{1}{4}\tilde{A}_{map}-\frac{1}{2}j-\frac{1}{2}l_{a}-\frac{1}{2}l_{b}+\nu^0_{a,b}$\tabularnewline
$\widetilde{CC}=\frac{1}{4}\tilde{A}_{map}+\frac{1}{2}j+\frac{1}{2}m_{a}+\frac{1}{2}m_{b}-\nu^0_{a,b}$\tabularnewline
$\widetilde{CD}=\frac{1}{4}\left(\tilde{B}_{map}-n\right)+\frac{1}{2}j+\frac{1}{2}m_{a}-\frac{1}{2}m_{b}+\nu^0_{a,b}$\tabularnewline
$\widetilde{DC}=\frac{1}{4}\left(\tilde{B}_{map}-n\right)+\frac{1}{2}j-\frac{1}{2}m_{a}+\frac{1}{2}m_{b}+\nu^0_{a,b}$\tabularnewline
$\widetilde{DD}=\frac{1}{4}\tilde{A}_{map}+\frac{1}{2}j-\frac{1}{2}m_{a}-\frac{1}{2}m_{b}-\nu^0_{a,b}$\tabularnewline
\end{tabular}
\par\end{centering}
\caption{Base-16 counts as functions of quantum numbers for the model of interest  \label{tab: base-8} }
\end{table}

\nocite{Powers_code_2023}

\bibliography{references}

\end{document}